\documentclass[lettersize,journal]{IEEEtran}
\usepackage{amsmath,amsfonts}
\usepackage{algorithmic}
\usepackage{algorithm}
\usepackage{array}
\usepackage[caption=false,font=normalsize,labelfont=sf,textfont=sf]{subfig}
\usepackage{textcomp}
\usepackage{stfloats}
\usepackage{url}
\usepackage{verbatim}
\usepackage{graphicx}
\usepackage{hyperref}
\newcolumntype{M}[1]{>{\centering\arraybackslash}m{#1}}
\usepackage{booktabs}
\usepackage{tikz}
\usetikzlibrary{positioning, fit, shapes, arrows.meta}
\usetikzlibrary{matrix, positioning, fit, shapes, arrows.meta,calc}
\usepackage{pifont}
\usepackage{multirow}
\usepackage{rotating}
\usepackage{makecell}
\usepackage{adjustbox}
\usepackage{xurl}
\usepackage{enumitem}
\usepackage{xcolor}
\usepackage{tikz}
\usetikzlibrary{positioning, arrows.meta, fit}
\usepackage{tikz}
\usetikzlibrary{positioning, arrows.meta, matrix, calc}
\usepackage{xcolor}

\definecolor{mainfill}{RGB}{225,236,248}   
\definecolor{maindraw}{RGB}{90,120,160}

\definecolor{intrafill}{RGB}{237,231,246}  
\definecolor{intradraw}{RGB}{126,103,165}

\definecolor{interfill}{RGB}{230,242,230}  
\definecolor{interdraw}{RGB}{102,140,102}

\definecolor{intraexfill}{RGB}{246,242,250} 
\definecolor{interexfill}{RGB}{241,248,241} 
\definecolor{mainexfill}{RGB}{240,246,252}  
\newcommand{\red}[1]{\textcolor{black}{#1}}

\newcommand{\redd}[1]{\textcolor{black}{#1}}
\usepackage{booktabs}
\usepackage{tabularx}

\begin{document}

\title{Agent System Operations: Categorization, Challenges, and Future Directions}

\author{
\IEEEauthorblockN{Zexin Wang\IEEEauthorrefmark{1}\IEEEauthorrefmark{6}, Changhua Pei\IEEEauthorrefmark{1}\IEEEauthorrefmark{2}\thanks{Corresponding author: Changhua Pei. Email: chpei@cnic.cn}}, Yuanhao Liu\IEEEauthorrefmark{2}, Jingjing Li\IEEEauthorrefmark{1}, Yintong Huo\IEEEauthorrefmark{5}, Quan Zhou\IEEEauthorrefmark{1}, Haotian Si\IEEEauthorrefmark{1}, Hang Cui\IEEEauthorrefmark{1}, Zihan Liu\IEEEauthorrefmark{1}, Jianhui Li\IEEEauthorrefmark{1}, Gaogang Xie\IEEEauthorrefmark{1},~\IEEEmembership{Senior Member,~IEEE}, Fei Sun\IEEEauthorrefmark{4}, Dan Pei\IEEEauthorrefmark{4},~\IEEEmembership{Senior Member,~IEEE},\\David Lo\IEEEauthorrefmark{5},~\IEEEmembership{Fellow,~IEEE} 


\IEEEauthorblockA{\IEEEauthorrefmark{1} Computer Network Information Center, Chinese Academy of Sciences} \\
\IEEEauthorblockA{\IEEEauthorrefmark{2} Hangzhou Institute for Advanced Study, University of Chinese Academy of Sciences}
\IEEEauthorblockA{\IEEEauthorrefmark{3} Tsinghua University} \\
\IEEEauthorblockA{\IEEEauthorrefmark{4} Institute of Computing Technology, Chinese Academy of Sciences}

\IEEEauthorblockA{\IEEEauthorrefmark{5} Singapore Management University}
\IEEEauthorblockA{\IEEEauthorrefmark{6} University of Chinese Academy of Sciences}
\thanks{This work was funded by the National Natural Science Foundation of China (62202445), and the National Natural Science Foundation of China-Research Grants Council (RGC) Joint Research Scheme (62321166652).}
}

\maketitle

\markboth{Accepted by IEEE Transactions of Software Engineering}%
{Shell \MakeLowercase{\textit{Wang et al.}}: Agent System Operations: Categorization, Challenges, and Future Directions}


\begin{abstract}
As the reasoning capabilities of Large Language Models (LLMs) continue to advance, LLM-based agent systems offer advantages in flexibility and interpretability over traditional systems, garnering increasing attention. However, despite widespread research interest in and industrial application of agent systems, these systems, like their traditional counterparts, frequently encounter anomalies. These anomalies lead to instability and insecurity, hindering their further development. Therefore, a comprehensive and systematic approach to the operation and maintenance of agent systems is urgently needed. Unfortunately, current research on the operations of agent systems is sparse. To address this gap, we have undertaken a survey on agent system operations with the aim of establishing a clear framework for the field, defining the challenges, and facilitating further development. Specifically, this paper begins by systematically defining anomalies within agent systems, categorizing them into intra-agent anomalies and inter-agent anomalies. Next, we introduce a novel and comprehensive operational framework for agent systems, dubbed \textbf{Agent} System \textbf{Op}eration\textbf{s} (\textbf{AgentOps}). We provide detailed definitions and explanations of its four key stages: monitoring, anomaly detection, root cause localization, and resolution.
\end{abstract}


\section{Introduction}

\IEEEpubidadjcol
\IEEEPARstart{W}{ith} the advent of technologies such as
DeepSeek-R1 \cite{guo2025deepseekr1} and Claude \cite{claude}, the reasoning capabilities of current Large Language Models (LLMs) are continually being enhanced. Leveraging LLMs as powerful cognitive engines, existing LLM-based agent systems, particularly multi-agent systems, have gained the capacity to accomplish a wide array of complex tasks and social simulations \cite{li2023econagent}, especially when equipped with diverse tools \cite{qin2023toolllm}. Compared to traditional systems like microservice architectures \cite{pei2025flow}, agent systems offer better automation, enhanced interpretability, and greater flexibility. Consequently, research and industrial applications of agent systems are flourishing, with an increasing number of online services \cite{online}, such as customer support and recommendation systems, adopting these agent systems.


However, despite the widespread application of agent systems, they are not without their flaws. Compared to traditional microservice systems, the greater flexibility offered by agent systems also introduces more anomalies \cite{wang2026trajaudit}. As illustrated in Fig.~\ref{fig:examples}, task execution often fails due to issues such as hallucinations. In role-playing scenarios, an attack on a single agent can lead to the collapse of the entire simulation. Therefore, to maintain the security and stability of agent systems and to facilitate their further development, efficient operations and maintenance are necessary.


\IEEEpubidadjcol


Operations technologies have evolved substantially over time, progressing from manual operations to rule-based approaches and, more recently, Artificial Intelligence for IT Operations (AIOps). However, agent systems exhibit fundamentally different characteristics and failure modes from traditional software systems, limiting the direct applicability of existing operations techniques. For instance, conventional log-based anomaly detection may identify abnormal patterns in system logs, but it is inherently inadequate for detecting agent-specific failures, such as orchestration hallucinations. These emerging challenges call for a new generation of operations technologies specifically designed for agent systems.


Currently, there is a lack of comprehensive research on effective operations and maintenance strategies specifically for agent systems. Most studies remain focused on isolated aspects of agent systems rather than addressing their overall operational challenges. For example, Durante et al. \cite{lifeifei} expound on agent paradigms and classifications; Chakraborty et al. \cite{hallucinationsurvey} delve into hallucinations in foundation models, covering their definition and detection methods; Deng et al. \cite{security} explore security issues in multi-agent systems, primarily covering external malicious attacks and categorizing threats into intra-execution security and interaction security. Shi et al. \cite{guiagents} provide detailed insights into security issues and evaluation methods for GUI agents.


To further advance the development of agent systems, this paper introduces the concept of \textbf{Agent} System \textbf{Op}eration\textbf{s} (\textbf{AgentOps}), a novel operations and maintenance framework specifically designed for agent systems. First, we provide a precise definition of anomalies within agent systems and offer a systematic classification, primarily dividing them into intra-agent and inter-agent anomalies. These two categories encompass the stages of pre-execution, execution, and post-execution in the agent system lifecycle. Additionally, drawing inspiration from traditional operation practices, we divide the operations and maintenance process for agent systems into \red{four} phases: monitoring, anomaly detection, root cause localization, and resolution. For each phase, we identify new challenges that arise within agent systems and propose detailed definitions and potential solutions.

\redd{Beyond reviewing existing techniques, this survey distills the stage-specific challenges of AgentOps into two overarching directions for future research: (1) timely and proactive AgentOps, which seeks to anticipate, detect, and mitigate failures before they propagate or escalate; (2) accurate and evidence-grounded AgentOps, which leverages rich, reliable, and context-aware operational evidence to support precise diagnosis and effective recovery. Together, these directions point toward a transition from today’s fragmented and predominantly post-hoc practices to proactive, evidence-grounded, and closed-loop operational paradigms for dependable agent systems.}

In section \ref{anomalies}, we present a detailed discussion of anomaly classification within agent systems. Section \ref{sec:4} introduces the concept of AgentOps. The subsequent four sections focus respectively on key components of AgentOps: monitoring, anomaly detection, root cause localization (also known as failure attribution \cite{icml2025}), and resolution. Section \ref{Datasets} introduces the datasets and benchmarks. Section \ref{CONCLUSION} concludes the paper.


\section{Anomalies in Agent Systems}
\label{anomalies}

\subsection{Definition of Anomalies in Agent Systems}


\redd{The HAL leaderboard \cite{hal} shows that success rates remain relatively low across different types of agent systems}, suggesting the presence of numerous exceptional cases that impede successful task completion. Prior work, such as \textit{Who\&When}~\cite{icml2025}, hypothesizes that anomalies in agent systems primarily arise at specific steps during task execution. Here, \emph{execution} refers to the phase in which an agent system, after receiving a user instruction, begins reasoning and invokes tools to carry out the task.

\begin{figure}[t]
    \centering
    \includegraphics[width=0.4\textwidth]{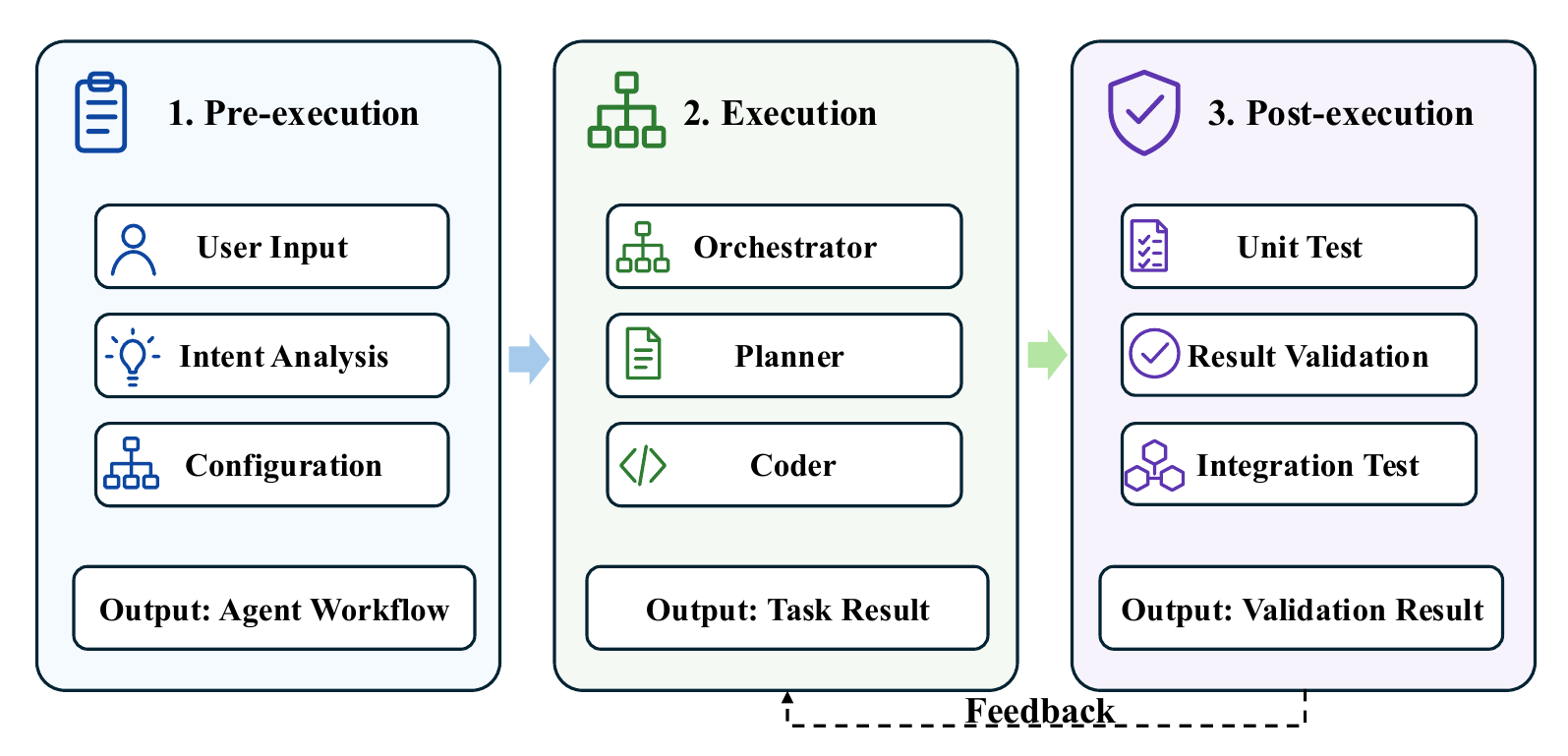}  
    \caption{\redd{Workflow of agent systems.}}
    \label{fig:flow}  
\end{figure}

\red{However, modern definitions of agent systems extend well beyond the execution phase alone. Both the pre-execution and post-execution phases are also critical to task success. 
First, in the \emph{pre-execution} phase, as illustrated in Fig.~\ref{fig:flow}, agent frameworks such as Claude Code, Hermes Agent, OpenClaw, and Deep Agent typically perform intent recognition based on the user input and subsequently configure or construct an agent workflow. Such workflows are not static; rather, they are dynamically instantiated and continuously updated as the task evolves. 
Second, in the \emph{post-execution} phase, task completion does not necessarily imply task success. For example, when Claude Code builds a software project, the process should not be considered successful merely because the build process terminates; the generated project may still contain bugs or fail to satisfy functional requirements. Similarly, in the domain of operations and maintenance, a root-cause analysis task may proceed without any apparent execution-time abnormality, yet still fail due to premature termination or insufficiently fine-grained analysis. These cases demonstrate that post-execution factors can also critically affect task outcomes. Therefore, we define anomalies in agent systems as any events occurring during the pre-execution, execution, or post-execution phases that disrupt task progression or prevent effective task completion.}

\begin{figure}[t]
    \centering
    \includegraphics[width=0.4\textwidth]{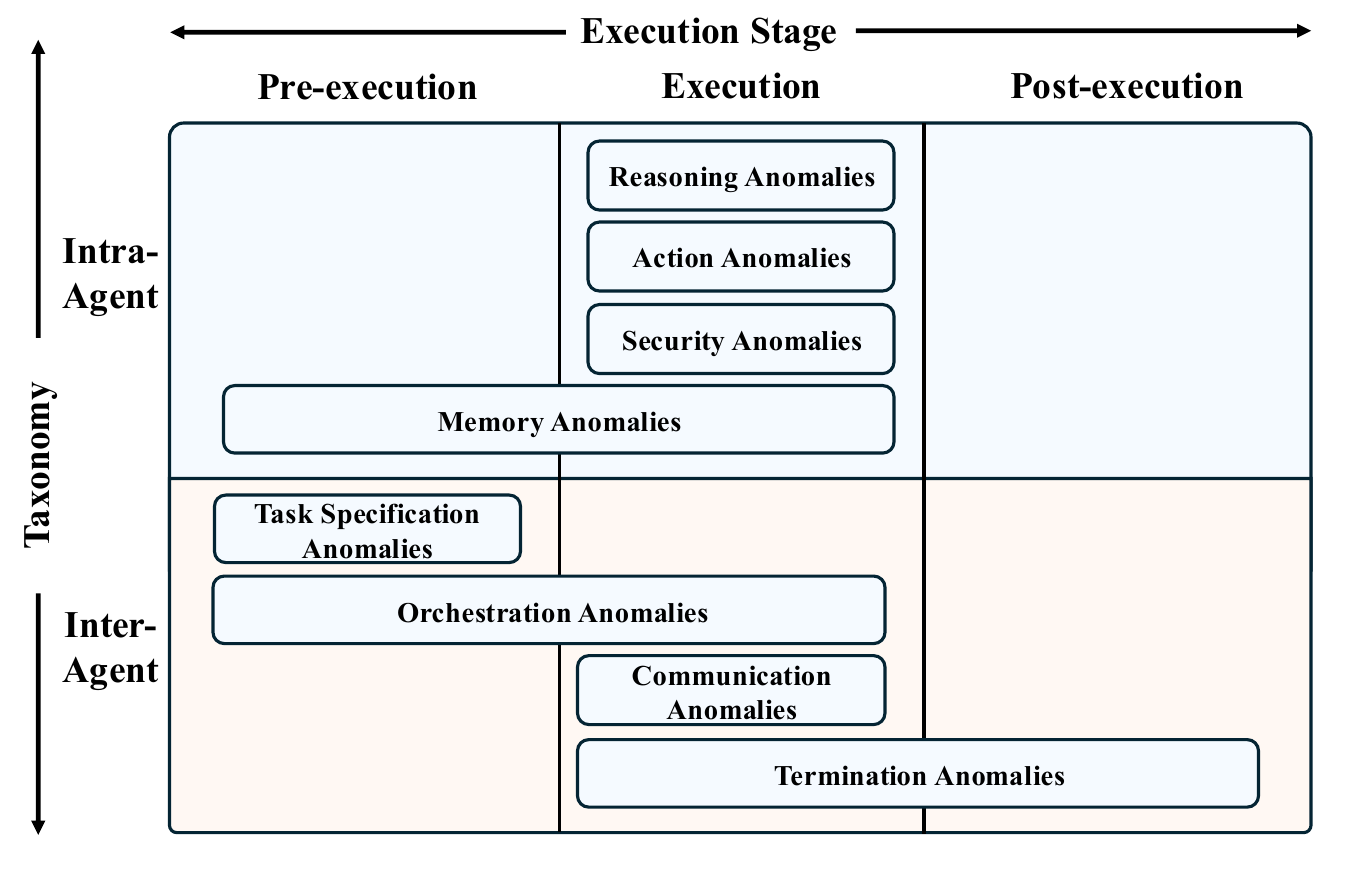}  
    \caption{\red{Definition and taxonomy of anomalies in agent systems.}}
    \label{fig:anomalies}  
\end{figure}


\red{Building on the above phase-aware definition, we further analyze anomalies from an orthogonal structural perspective. 
The pre-execution, execution, and post-execution phases clarify \emph{when} anomalies may occur, but they do not specify \emph{where} such anomalies originate within an agent system. 
Since agent systems can be broadly classified into single-agent and multi-agent systems, anomalies may originate either from the internal workflow of an individual agent or from interactions among multiple agents. 
This is analogous to traditional service architectures, where failures may arise within a single service or during inter-service communication. 
Therefore, as depicted in Fig.~\ref{fig:anomalies}, we categorize anomalies into two major types: \emph{intra-agent anomalies} and \emph{inter-agent anomalies}. 
Intra-agent anomalies refer to disruptions rooted in an individual agent's reasoning or tool use, while inter-agent anomalies refer to failures caused by coordination, communication, or dependency issues among agents.}

\subsection{Intra-Agent Anomalies}

Intra-agent anomalies refer to failures that arise within an individual agent or subagent during task execution. Since most existing agent implementations follow the ReAct paradigm~\cite{react}, such anomalies commonly manifest as reasoning and action failures. In addition, when agents are deployed in complex vertical domains, naive designs often suffer from memory degradation and security risks caused by long-context processing, unreliable retrieval, and insufficiently constrained tool permissions. These observations highlight the importance of harness engineering for improving the robustness, controllability, and safety of agent systems.

\subsubsection{Reasoning Anomalies}

Reasoning is the cognitive foundation that guides an agent's subsequent actions and enables complex task completion. Recent studies have proposed various techniques to enhance reasoning, including fine-tuning methods such as SFT~\cite{sft} and DeepSeek-R1~\cite{guo2025deepseekr1}, as well as prompting strategies such as CoT~\cite{wei2022chainofthought}. Despite these advances, reasoning anomalies remain prevalent, with hallucination being the most representative example.

The definition of hallucination has been continuously refined in the literature. Rawte et al.~\cite{109} define hallucinations as unreliable generations that contradict known facts, while Gallifant et al.~\cite{39} regard them as responses unrelated to the original prompt. Chakraborty et al.~\cite{hallucinationsurvey} summarize hallucinations through four characteristics: compliance, desirability, relevancy, and plausibility. Yang et al.~\cite{yang2024alignment} further interpret hallucination as a form of dishonesty, where uncertain answers are expressed with unwarranted confidence. In general, hallucinations can be viewed as anomalous generations that deviate from facts, logic, or task requirements. They are difficult to eliminate because LLMs are sensitive to training data, prone to knowledge forgetting, and unable to synchronously incorporate newly emerging knowledge after training.

\subsubsection{Action Anomalies}

Action anomalies occur when agents fail to correctly execute intended operations through tool invocation or function calls. In practice, such failures may arise from inconsistent interfaces, incorrect API selection, delayed execution, malformed parameters, or system-level failures~\cite{industrialfunction}. Moreover, tool use introduces additional safety risks. Wu et al.~\cite{wu2024darkside} show that attackers can exploit function-calling mechanisms through jailbreak prompts, inducing LLMs to invoke sensitive functions or bypass restrictions. Although the Model Context Protocol (MCP) standardizes interactions between LLMs and external tools, it does not fully eliminate action anomalies. In real deployments, configuration changes and misconfigured MCP servers may still lead to failed, unintended, or unsafe actions~\cite{ai-infra-guard}.

\subsubsection{Memory Anomalies}

Agent memory can be broadly divided into short-term and long-term memory. Short-term memory corresponds to the LLM context. Although the context windows of LLMs continue to expand, they often remain insufficient for complex tasks. Many agent frameworks therefore adopt sliding-window strategies, which may discard early but important instructions. Even when the context length is technically sufficient, LLMs may still underutilize information located in the middle of long contexts, a phenomenon known as the ``lost-in-the-middle'' problem~\cite{liu2023lostinmiddle}. PI-LLM~\cite{PI-LLM} also shows that LLMs face intrinsic bottlenecks in working memory.

Long-term memory is typically implemented through vector databases and recalled using RAG. However, retrieval-based memory is vulnerable to inaccurate recall, noisy evidence, and unreliable generation. QE-RAG~\cite{zhang2025qerag} shows that existing RAG systems are highly sensitive to noise, while Astute RAG~\cite{wang2024astuterag} highlights conflicts between internal parametric knowledge and retrieved external knowledge. Such conflicts may further induce RAG hallucinations~\cite{sun2024redeep}. Chen et al.~\cite{chen2024benchmarkrag} also observe that despite extensive technical improvements, the accuracy of current RAG systems remains limited.

\subsubsection{Security Anomalies}

Security anomalies refer to cases in which an agent or subagent performs operations that violate predefined security policies, access-control rules, or identity constraints. Typical examples include unauthorized tool invocation, actions beyond assigned privileges, misuse of delegated credentials, and operations executed under an incorrect or unverified identity. Since LLM-based agents interact with tools and APIs through natural-language reasoning, their security boundaries are often implicit and difficult to enforce.

Importantly, security anomalies do not necessarily originate from an agent's autonomous intention or internal malfunction. They may be induced by adversarial inputs, prompt injection, malicious tool outputs, or compromised external environments, which manipulate the agent into actions outside its intended authorization scope. These risks are closely related to excessive agency, where agents are granted more autonomy or privileges than necessary, allowing erroneous or manipulated outputs to trigger unsafe operations~\cite{blogemergent,zhang2025agent}. In multi-agent systems, security anomalies can further arise from improper authorization propagation during delegation or inter-agent communication, where a compromised agent indirectly causes other agents to perform unauthorized actions. Therefore, robust agent systems require explicit identity management, least-privilege tool access, adversarially robust input and tool-output handling, and policy-aware execution mechanisms.

\subsection{Inter-Agent Anomalies}

Inter-agent anomalies refer to failures emerging from the interaction, coordination, and collective behavior of multiple agents. Unlike fixed workflows, many recent frameworks, such as Hermes, dynamically construct subagents from available skills and tools according to user inputs, then perform planning, execution, and result verification. As a result, inter-agent anomalies may occur throughout the entire lifecycle. In the pre-execution stage, they mainly involve task specification and orchestration failures caused by ambiguous, incomplete, or malicious inputs and inappropriate subagent construction. During execution, communication anomalies may arise when agents exchange information, invoke tools, or rely on unreliable intermediate outputs. In the post-execution stage, termination anomalies occur when the system incorrectly determines task completion or result validity.


\subsubsection{Task Specification Anomalies}

Task specification anomalies arise when user goals, task constraints, agent roles, or configurations are ambiguous, incomplete, or inconsistent. Pan et al.~\cite{whydomasfail} show that many task-level failures originate from unclear task definitions, such as insufficiently specified prompts. Altmann et al.~\cite{task1} further observe that poorly defined tasks may lead to chasing and blocking, even when each agent's local behavior appears reasonable. SentinelAgent~\cite{he2025sentinelagent} reports that insufficiently specified collaboration modes may cause agents to deviate from intended behaviors, collude, or become vulnerable to prompt injection. Therefore, evaluating task completeness before execution and enabling reflection during execution are important for robust multi-agent operation.

Agent configuration is also an essential component of task specification, including prompts, roles, skills, and tools. Pan et al.~\cite{whydomasfail} identify incorrect role configuration as a common source of anomalies. Platon et al.~\cite{conf1} show that unclear configurations and role confusion frequently lead to coordination failures and adversarial misalignment. Similarly, OG~\cite{confblog} argues that actions outside an agent's designated responsibility can directly cause conflicts, inconsistencies, and inefficiency, while AgentFM~\cite{zhang2025agentfm} identifies vague role configuration as a common failure source for database agents.

\subsubsection{Orchestration Anomalies}

\red{Orchestration anomalies refer to failures in the global planning process of agent systems. Modern LLM-based multi-agent systems usually involve task decomposition, role assignment, tool and skill selection, inter-agent coordination, and workflow control~\cite{li2024survey,wu2024autogen,li2023camel}. Errors at this stage may propagate to later execution even before any individual subagent takes action. Typical orchestration anomalies include missing or redundant subtasks, mismatched agent--tool allocation, circular or conflicting dependencies, inefficient workflows, and plans that violate task constraints or security policies. These anomalies are challenging because they are often not attributable to a single faulty agent, but instead emerge from improper global coordination, incomplete task understanding, or unstable planning decisions.}

\subsubsection{Communication Anomalies}

Communication anomalies occur during message exchange among agents. Bronsdon~\cite{blog} highlights message storms as a typical failure mode, where excessive inter-agent communication causes resource exhaustion, increased latency, and eventual task failure. AgentPrune~\cite{zhang2024agentprune} similarly observes that redundant messages do not necessarily improve system performance; instead, excessive communication may distract agents and reduce coordination efficiency. Therefore, effective communication control is essential for scalable and reliable multi-agent systems.

\subsubsection{Termination Anomalies}

Termination anomalies occur when an agent system incorrectly determines whether a task should stop or continue. Premature termination is a representative case, where the system stops before completing the required reasoning or execution steps. Pan et al.~\cite{whydomasfail} and Microsoft~\cite{baipishu} both regard premature stopping as an important category of agent-system failures. Smurfs~\cite{chen2025smurfs} shows that in single-agent settings, DFSDT may invoke the termination tool too early, thereby harming the completeness and logical coherence of complex tasks. Zhang et al.~\cite{icml2025} further identify premature termination as a common and locatable failure source in multi-agent systems.

In contrast, agents may also fail to terminate. Zhu et al.~\cite{recur} identify undercommitment anomalies, where an agent repeatedly delegates tasks to subagents, forming an infinite recursion chain that eventually reaches recursion-depth limits or timeouts. Drake~\cite{recur2} describes a related phenomenon, termed ``neural howlround,'' in which agents become trapped in recursive self-optimization or reasoning loops without a clear endpoint. Such non-termination anomalies may stall the system and lead to severe inefficiency or cognitive stagnation.

\section{What is AgentOps?}
\label{sec:4}

\begin{figure*}[t]
    \centering
    \includegraphics[width=0.82\textwidth]{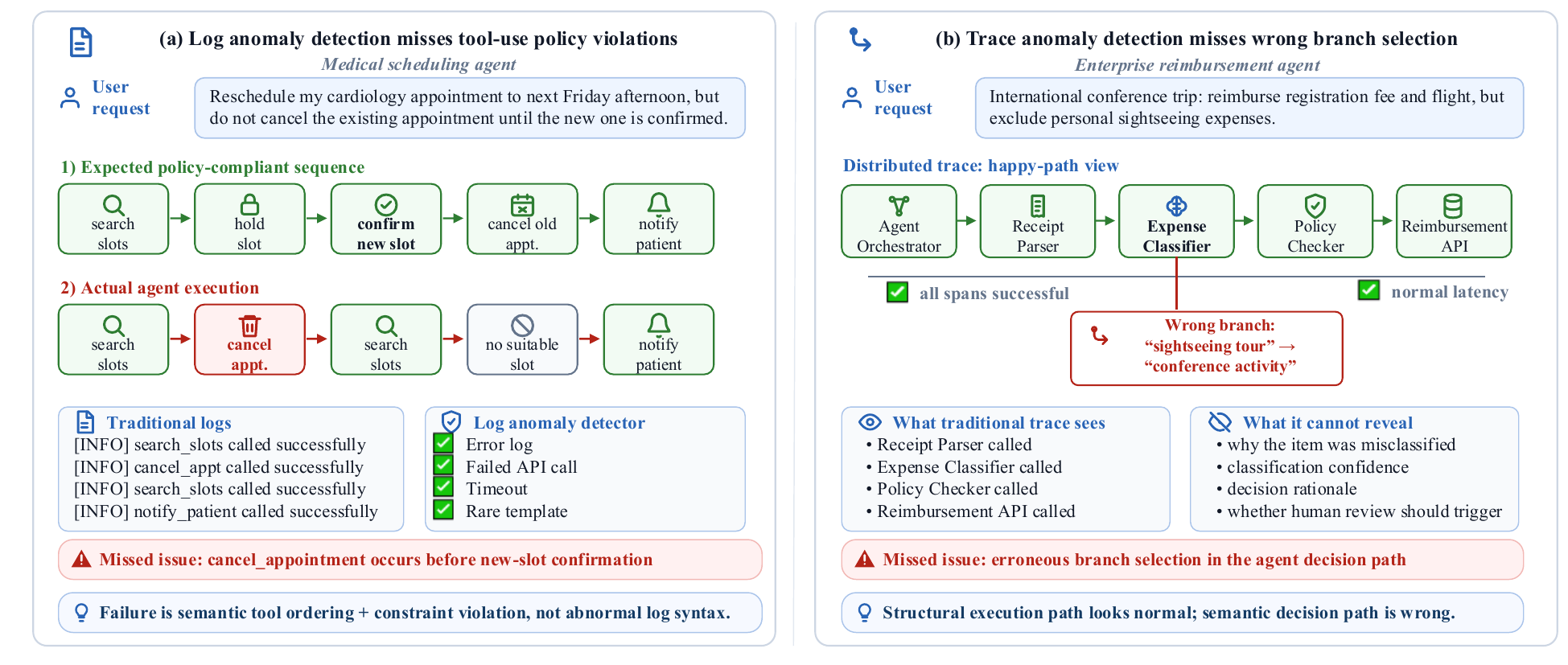}  
    \caption{\redd{Examples of applying traditional operations methods to agent systems. Traditional log-based anomaly detection (left) and trace-based anomaly detection (right) fail to detect anomalies in agent systems.}}
    \label{fig:examples}  
\end{figure*}




To start, Table \ref{tab:definition} presents key concepts that will be referenced throughout the discussion.

\begin{table}[t]
\centering
\caption{Definition of some concepts in the paper.}
\label{tab:definition}
\footnotesize
{
\begin{tabular}{>{\raggedright\arraybackslash}m{2cm} p{6cm}}
\toprule
\textbf{Name}     & \multicolumn{1}{c}{\textbf{Definition}}                                                                                                                                                           \\ 
\midrule
\textbf{DevOps} \cite{smeds2015devops}     & DevOps merges software development and IT operations to shorten the development lifecycle and deliver high-quality software.
     \\
\midrule
\textbf{AIOps} \cite{cheng2023ai}   & AIOps uses machine learning to automate IT operations.                                                              \\
\midrule

\textbf{\redd{LLMOps}} \cite{llmops} & \redd{LLMOps extends MLOps to support the deployment, monitoring, and optimization of LLM-based applications. AgentOps focuses on agent systems.} \\

\midrule
\textbf{AgenticOps} \cite{cisco2026agenticops} & Agentic Operations (AgenticOps) refers to the use of agents for the maintenance of traditional systems. \\
\midrule
\textbf{AgentOps} & AgentOps refers to the use of various operational techniques for the maintenance of agent systems.  \\
\bottomrule
\end{tabular}
}
\end{table}

\subsection{Evolution of Operations}

\begin{figure}[t]
    \centering
    \includegraphics[width=0.49\textwidth]{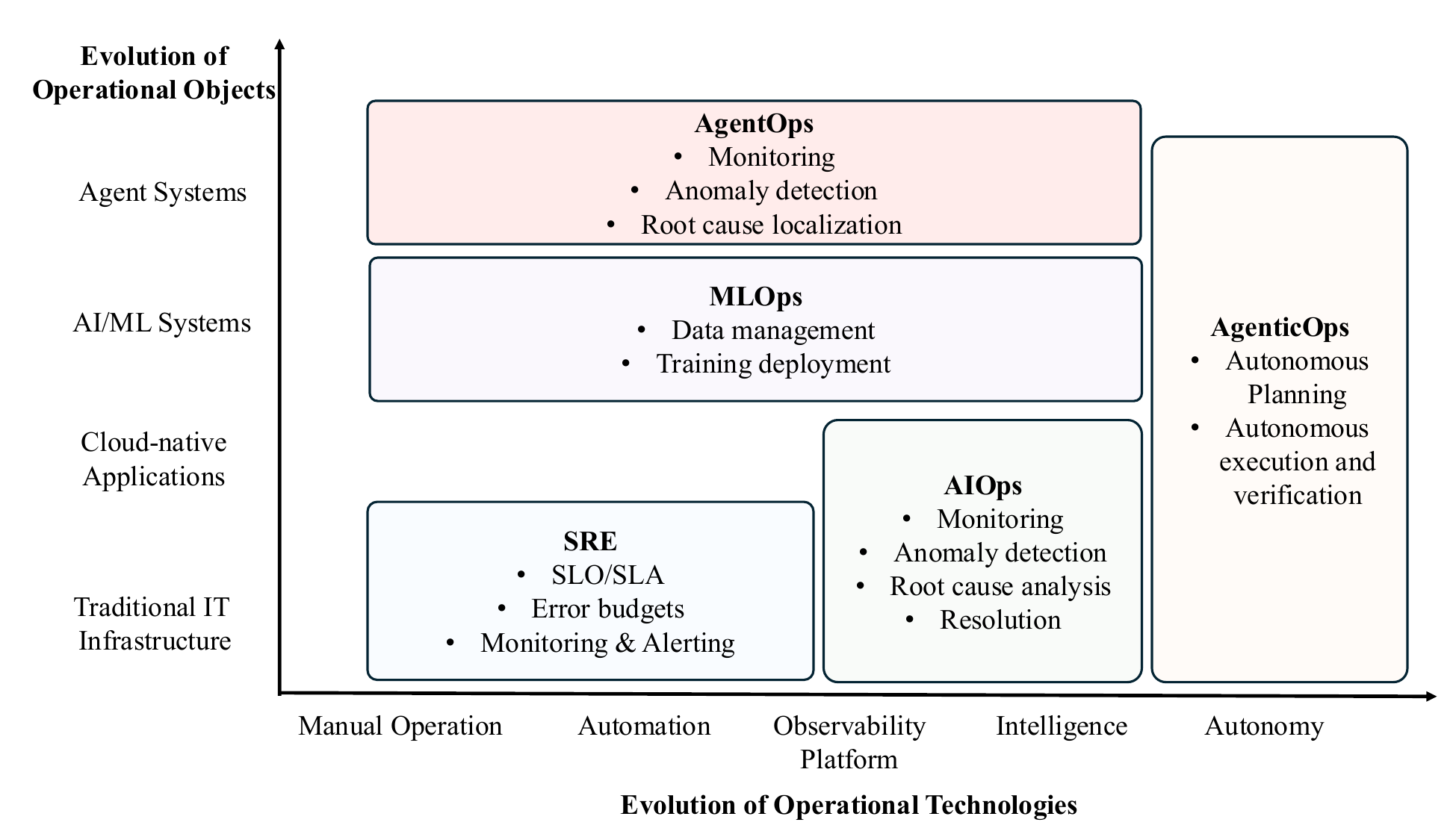}  
    \caption{\redd{Evolution of operational objects and operational technologies over time. SLO denotes Service Level Objective, while SLA denotes Service Level Agreement.}}
    \label{fig:agentopsandopsagent}  
\end{figure}

\red{As shown in Fig.~\ref{fig:agentopsandopsagent}, the evolution of operations can be understood from two dimensions: the evolution of operational technologies and the evolution of operational objects. Along the technology dimension, operations have advanced from manual intervention and rule-based automation to platform-based observability, intelligent analysis, and, more recently, autonomous operations enabled by LLM agents. Along the object dimension, the target of operations has expanded from traditional IT infrastructure and cloud-native applications to AI/ML systems and emerging agent systems.}

\red{Different operational paradigms correspond to different stages in this two-dimensional evolution. Site Reliability Engineering (SRE) focuses on ensuring the reliability of applications and infrastructure through engineering practices such as Service Level Objective (SLO) management, monitoring, alerting, and change control. AIOps further introduces intelligent techniques to maintain the stability of diverse systems, including anomaly detection, root cause analysis, alert noise reduction, and predictive optimization. As AI/ML systems became increasingly complex, MLOps emerged to manage the lifecycle and reliability of data pipelines, model training, deployment, inference, and drift monitoring.}

\red{Recently, LLM-based agents have become a new class of operational objects. Unlike traditional systems, whose behaviors are largely deterministic and specified by program logic, agent systems rely on probabilistic foundation models, dynamic tool invocation, memory, and multi-step reasoning. Their failures may arise not only from infrastructure faults or model degradation, but also from reasoning errors, unsafe actions, tool misuse, and misalignment with user or system goals. These characteristics make it insufficient to directly apply existing operational techniques to agent systems.}

This insufficiency can be illustrated by two representative failures. 
As shown in Fig.~\ref{fig:examples}, in a medical scheduling agent, the user asks to reschedule an appointment while keeping the original one until a new slot is confirmed. 
Although the correct trajectory should call \texttt{cancel\_appointment} only after \texttt{confirm\_reschedule}, the agent may prematurely cancel the existing appointment and later fail to find a new slot. 
Traditional log anomaly detection may not flag this behavior, since all tool calls can return successful status codes. 
The failure is instead a semantic tool-use policy violation: the agent violates the required ordering constraint. Similarly, in an enterprise reimbursement agent, a conventional trace may show a normal path from receipt parsing to expense classification, policy checking, and reimbursement submission, with all spans completing successfully. 
However, the agent may classify a ``sightseeing tour'' as a conference activity and reimburse it. 
The trace captures the structural service-call path, but not the semantic decision path behind the incorrect classification. 
These examples show that agent failures may be trajectory-level and semantic, involving intent, constraints, tool-use order, and decision rationale, rather than conventional anomalies in logs, status codes, latency, or traces.

\red{Therefore, we propose \textbf{AgentOps}, which refers to operational techniques for maintaining the reliability, safety, and controllability of agent systems. AgentOps is distinct from \textbf{AgenticOps}: the former operates \emph{on} agent systems, whereas the latter uses agent systems as operational actors to maintain other systems. This distinction motivates the need for a dedicated operational framework for agent systems.}

\subsection{Difference between Traditional Operations and AgentOps}
\label{sec:diff}


\begin{figure}[t]
    \centering
    \includegraphics[width=0.5\textwidth]{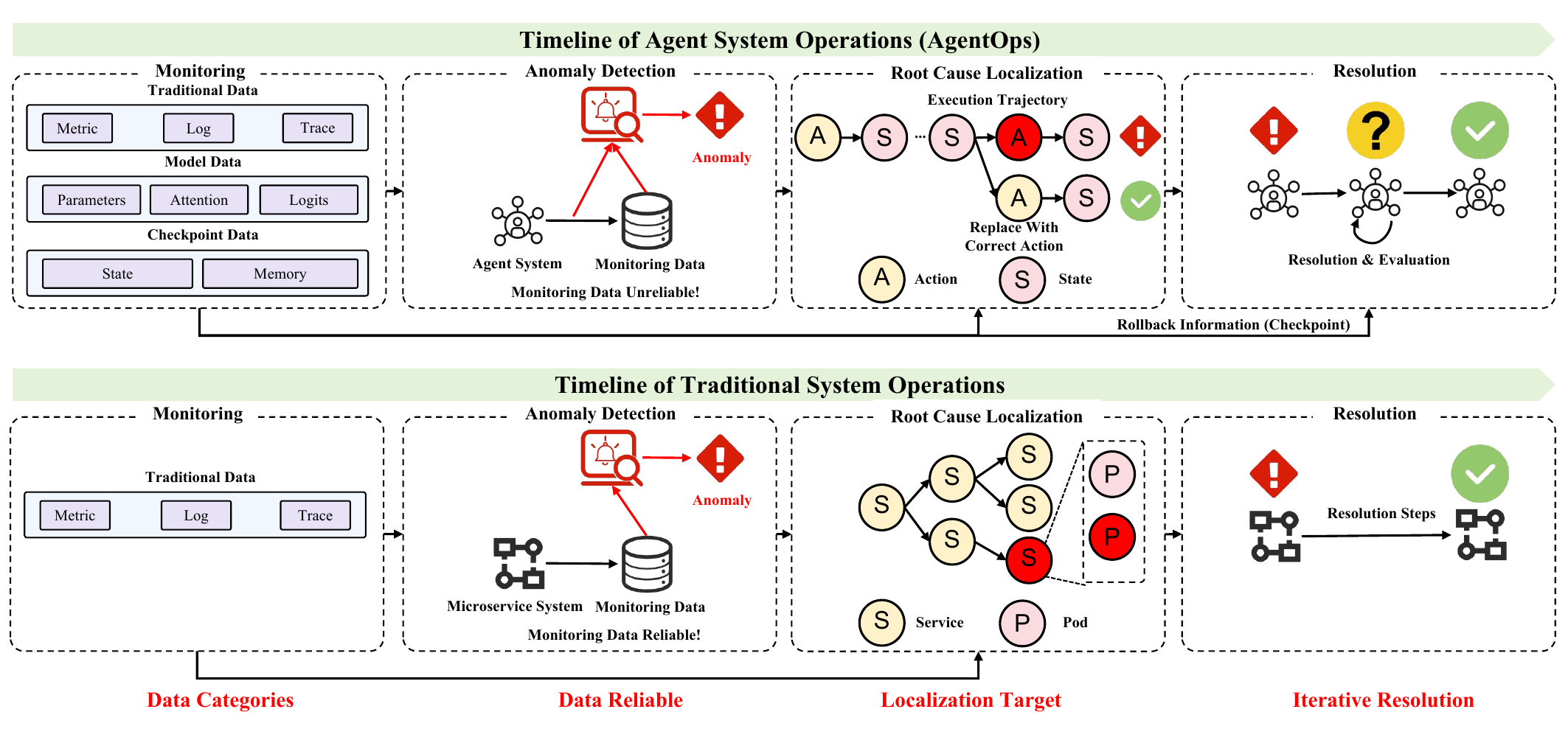}  
    \caption{Comparison of traditional system operations and agent system operations across four key stages.}
    \label{fig:operations}  
\end{figure}

\begin{table}[t]
\centering
\scriptsize
\setlength{\tabcolsep}{3pt}
\renewcommand{\arraystretch}{0.88}
\caption{\red{Task-Level mapping of reliability practices in DevOps and AgentOps.}}
\label{tab:practice_task_mapping}
\begin{tabularx}{\columnwidth}{
p{1.55cm}
p{1.8cm}
>{\raggedright\arraybackslash}X
>{\raggedright\arraybackslash}X
}
\toprule
\textbf{Stage} & \textbf{Task} & \textbf{DevOps} & \textbf{AgentOps} \\



\midrule

\multirow{3}{=}{\textbf{Anomaly detection}}
& \textbf{Process validation}
& Detects sequential breaks in text logs.
& Detects uncertainty or hallucination signals in outputs. \\

\cmidrule(lr){2-4}
& \textbf{Health monitoring}
& Finds anomalies in infrastructure metric streams.
& Finds loops, abnormal actions, or trajectory deviations. \\

\cmidrule(lr){2-4}
& \textbf{Output validation}
& Checks formats and deterministic thresholds.
& Uses LLM-as-a-judge or fact-checking for semantics. \\

\midrule

\textbf{Root cause localization}
& \textbf{Causal analysis}
& Localizes faults through static dependency graphs.
& Localizes flawed actions or reasoning steps via trajectories. \\

\midrule

\multirow{2}{=}{\textbf{Resolution}}
& \textbf{System recovery}
& Executes predefined runbooks without re-evaluation.
& Uses self-correction loops and prompt refinement. \\

\cmidrule(lr){2-4}
& \textbf{State rollback}
& Rolls back static database snapshots.
& Restores agent state to previous stable checkpoints. \\

\bottomrule
\end{tabularx}
\end{table}

As shown in Fig.~\ref{fig:operations}, although traditional system operations and agent system operations follow a similar timeline, they differ substantially at each stage. Table \ref{tab:practice_task_mapping} further provides a task-level mapping of reliability practices in DevOps and AgentOps. These differences motivate the need for a dedicated operations framework, AgentOps, as detailed below.

\begin{itemize}[leftmargin=*]
    \item \textbf{Monitoring}: The primary difference lies in the monitored data. Traditional system operations, typically based on OpenTelemetry \cite{opentelemetry2019}, focus on metrics, logs, and traces that reflect system runtime states. Agent systems require additional monitoring because their services are driven by inherently stochastic LLM agents. For the LLM component, relevant signals include model parameters, attention maps, token logits, and other internal states; for the agent component, checkpoints such as memory and environment states should be recorded at each step. Such agent-level monitoring improves system observability and enables rollback, providing stronger operational flexibility than traditional systems.

    \item \textbf{Anomaly Detection}: \red{The main difference lies in where anomalies may originate. In traditional systems, traces, logs, and metrics are generally generated by deterministic or rule-based mechanisms and can therefore serve directly as reliable inputs for anomaly detection. In agent systems, however, monitoring data may be produced or mediated by probabilistic LLM-based agents, whose reasoning and generation processes can themselves introduce anomalies. Consequently, anomaly detection must examine not only the observed monitoring data but also the underlying generation process. This motivates the use of model-level signals, such as attention maps and logits, to detect reasoning or generation anomalies, as discussed in section~\ref{sec:ad}.}

    \item \textbf{Root Cause Localization}: The primary distinction lies in the localization space and diagnosis target. Traditional operations localize faults over microservice dependency graphs, targeting infrastructure-level entities such as pods, services, or nodes. In contrast, agent systems localize failures along agent execution trajectories, where individual agent actions become the diagnosis targets. Accordingly, root cause localization in agent systems is also commonly referred to as failure attribution \cite{icml2025}.

    \item \textbf{Resolution}: The primary distinction lies in the resolution schema. In traditional systems, once the fault location and root cause are identified, remediation can often follow deterministic procedures based on domain knowledge. In agent systems, inherent stochasticity makes resolution more complex, often requiring iterative testing, refinement, and rollback based on monitoring data. \red{Moreover, traditional operations primarily adopt post-execution remediation, whereas AgentOps increasingly emphasizes intervention before and during execution. Because LLM-based agent systems may fail frequently and unpredictably, relying solely on reactive remediation can impose substantial operational overhead. Therefore, emerging approaches aim to prevent failures before execution or correct them in a timely manner during execution.}
\end{itemize}

\begin{figure*}[t]
    \centering
    \includegraphics[width=0.82\textwidth]{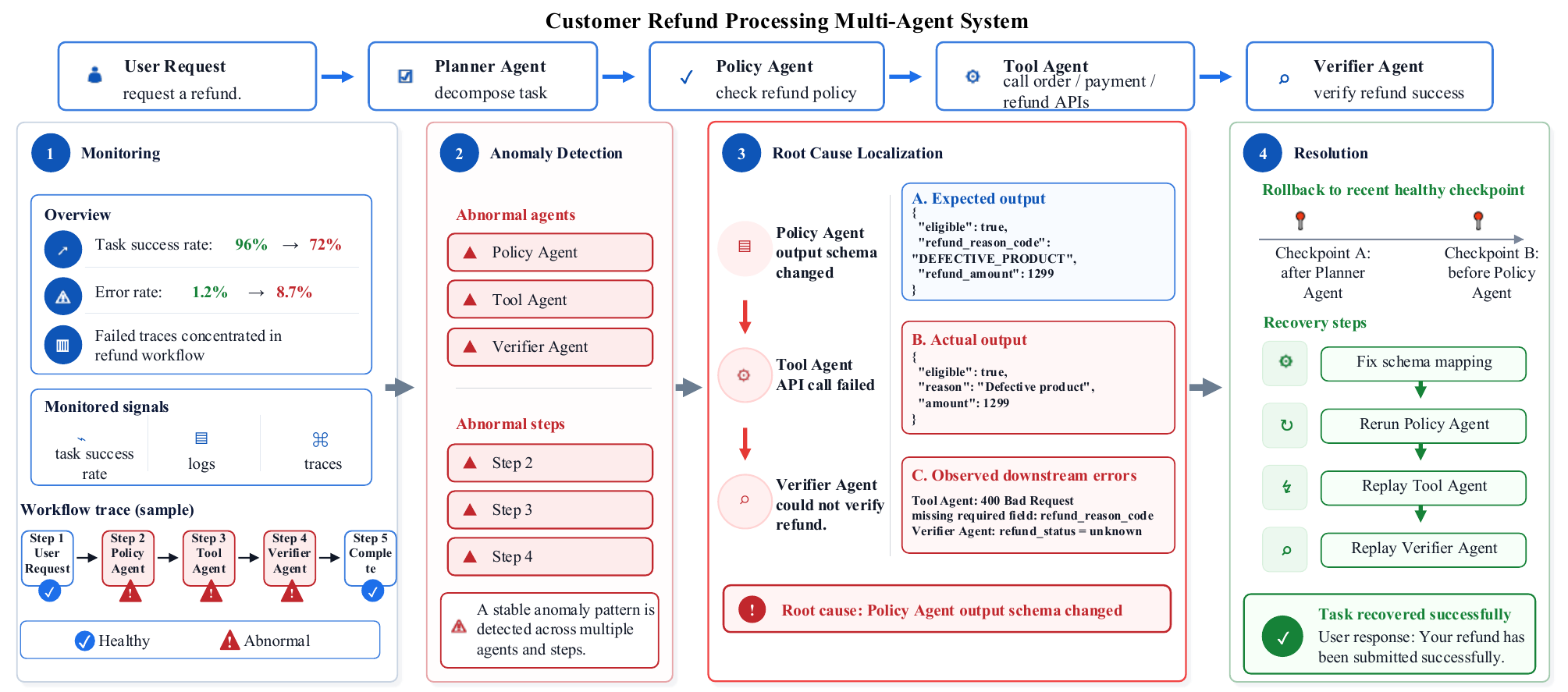}  
    \caption{\redd{Illustration of AgentOps for agent systems. AgentOps first detects anomalies and identifies the responsible steps and agents within the agent system, then performs root cause localization to pinpoint the underlying causes, and finally leverages checkpoint-based rollback for recovery.}}
    \label{fig:agentops-example}  
\end{figure*}

\subsection{Difference from Previous AgentOps Work}

\red{Existing literature on AgentOps generally approaches the domain through the lenses of observability, architectural adaptation, and role-based automation. For instance, Dong et al.~\cite{dong2024agentops} focus on establishing tracking and tracing pipelines for agent artifacts to ensure system observability. Meanwhile, Biswas et al.~\cite{Biswas_Bhatt_Vaidhyanathan_2026} and Moshkovich et al.~\cite{Moshkovich_Zeltyn_2025} focus on the dynamic nature of agents, proposing frameworks, such as CHANGE and the AgentOps Automation Pipeline, to manage continuous learning, multi-agent consensus, and role-specific optimization.}

\red{In contrast to prior work, this study conceptualizes AgentOps as a structured lifecycle for failure diagnostics and recovery. The distinction is twofold: (1) We shift the focus from traditional metric-based tracing to a taxonomy rooted in agent structural boundaries, encompassing both intra-agent and inter-agent dimensions. (2) Bridging the gap between high-level recommendations and low-level control, our framework introduces concrete execution-level interventions; by utilizing checkpointing and state rollbacks, it effectively addresses the stochastic nature of agent execution paths.}



\subsection{\redd{Rationale for the Four-Stage Design}}



Inspired by the differences discussed above, we define \textbf{Agent} System \textbf{Op}eration\textbf{s} (\textbf{AgentOps}) as a comprehensive operational framework that spans the pre-execution, execution, and post-execution stages of agent systems. Similar to traditional system operations, AgentOps consists of four key phases: monitoring, anomaly detection, root cause localization, and resolution. \red{Fig.~\ref{fig:agentops-example} illustrates these phases through a customer refund processing example, where multiple agents collaboratively decompose a refund request, check refund eligibility, invoke order/payment/refund APIs, and verify the final refund status. When a schema change in the Policy Agent propagates downstream and causes failures in the Tool Agent and Verifier Agent, AgentOps first detects the abnormal patterns from metrics, logs, and traces, then localizes the true upstream root cause, and finally restores the workflow by rolling back to a healthy checkpoint and replaying the affected steps. We introduce each phase in the following sections.}



\redd{To test whether each phase of AgentOps is useful, we create small workflows that resemble familiar online services. Consider a refund task: an automated program receives a refund request, reads the refund policy and two order records, and then uses tools to select and update the requested order. This tool-using program is the agent in our experiment. In a faulty run, we deliberately alter one decision---for example, the program selects the other order, skips the policy check, or uses an outdated observation. These mistakes represent reasoning or action anomalies and may affect later operations. Moreover, every order update leaves a permanent audit record. The program may subsequently correct the final order status, but it cannot erase the fact that it previously modified the wrong order. We therefore evaluate both whether the final result is correct (\emph{functional correctness}) and whether the workflow avoided unsafe operations (\emph{operational safety}); \emph{composite pass} requires both conditions.}

\redd{The benchmark contains 32 tasks across configuration management, order processing, incident recovery, and data publication (similar to Fig. \ref{fig:agentops-example}). Each task is repeated three times, producing 96 paired samples. We compare four cumulative settings: (1) a naive agent receives no intervention; (2) resolution is triggered only after the final evaluator reports a binary failure; (3) anomaly detection monitors the complete trajectory and interrupts execution when a high-confidence anomaly is observed; and (4) root cause localization further identifies the earliest causal step and provides this evidence to the resolver. All settings reuse the same trajectory prefix, environment state, and original agent. The evaluator never reveals the expected state or a failure diff. We use \texttt{deepseek-v4-flash} for the agent and operational modules, with thinking mode disabled. Monitoring includes assistant messages, tool calls, tool observations, errors, and state snapshots.}


\begin{table}[t]
    \centering
    \caption{\redd{Cumulative evaluation of the AgentOps phases over 96 paired samples. Pass and safety values are percentages; Steps is the mean number of tool steps.}}
    \label{tab:agentops-ablation}
    \setlength{\tabcolsep}{3.2pt}
    \begin{tabular}{@{}lcccc@{}}
        \toprule
        \textbf{Setting} & \textbf{Composite} & \textbf{Functional} & \textbf{Safety} & \textbf{Steps} \\
        \midrule
        \textbf{Naive} & 25.0 & 25.0 & 50.0 & 8.00 \\
        \textbf{+ Resolution} & 50.0 & 99.0 & 50.0 & 16.11 \\
        \textbf{+ Anomaly detection} & 85.4 & 97.9 & 87.5 & 12.48 \\
        \textbf{+ Root cause localization} & \textbf{87.5} & \textbf{100.0} & \textbf{87.5} & \textbf{11.24} \\
        \bottomrule
    \end{tabular}
\end{table}

\redd{Table~\ref{tab:agentops-ablation} demonstrates the distinct contribution of each phase. Monitoring is a passive prerequisite rather than a repair mechanism: it preserves both the agent's stated rationale and the resulting system behavior, enabling semantic inconsistencies to be distinguished from tool failures. With this evidence, anomaly detection identified all 72 faulty trajectories and did not intervene in any of the 24 clean trajectories in the controlled benchmark. Because monitoring is shared by all settings, however, its marginal contribution is not independently estimated here.}

\redd{Resolution alone doubled composite pass from 25.0\% to 50.0\% and increased functional correctness to 99.0\% (exact McNemar $p=1.19\times10^{-7}$). Safety nevertheless remained at 50.0\%, showing that post-execution repair can correct state but cannot undo an irreversible side effect. Adding anomaly detection increased composite pass to 85.4\% and safety to 87.5\% ($p=5.40\times10^{-9}$), while reducing the mean execution length by 3.64 steps relative to terminal resolution. The improvement arises because the agent is interrupted before an upstream anomaly propagates into downstream actions. Thus, anomaly detection contributes not merely another feedback message, but the timing required to prevent error propagation.}

\redd{Root cause localization yielded a smaller yet consistent incremental benefit. It repaired two additional samples without regressions and reduced execution by a further 1.24 steps. The pass-rate difference was not statistically significant ($p=0.50$), so we do not claim a general reliability improvement. Nevertheless, the localizer identified the exact causal step in all 72 faulty trajectories, and the reduced recovery length indicates that localized evidence helped the agent avoid unnecessary diagnostic actions. Taken together, the results support the proposed division of responsibilities: monitoring provides evidence, anomaly detection determines when to intervene, root cause localization narrows the repair target, and resolution executes the recovery. The remaining failures occurred when a harmful action was itself irreversible before the next monitoring checkpoint, highlighting a boundary of checkpoint-based operation rather than implying complete fault coverage.}

\section{How to Monitor Agent Systems?}
\label{sec:monitor}

\subsection{Illustration of Monitoring Data}


\begin{figure}[t]
    \centering
    \includegraphics[width=0.5\textwidth]{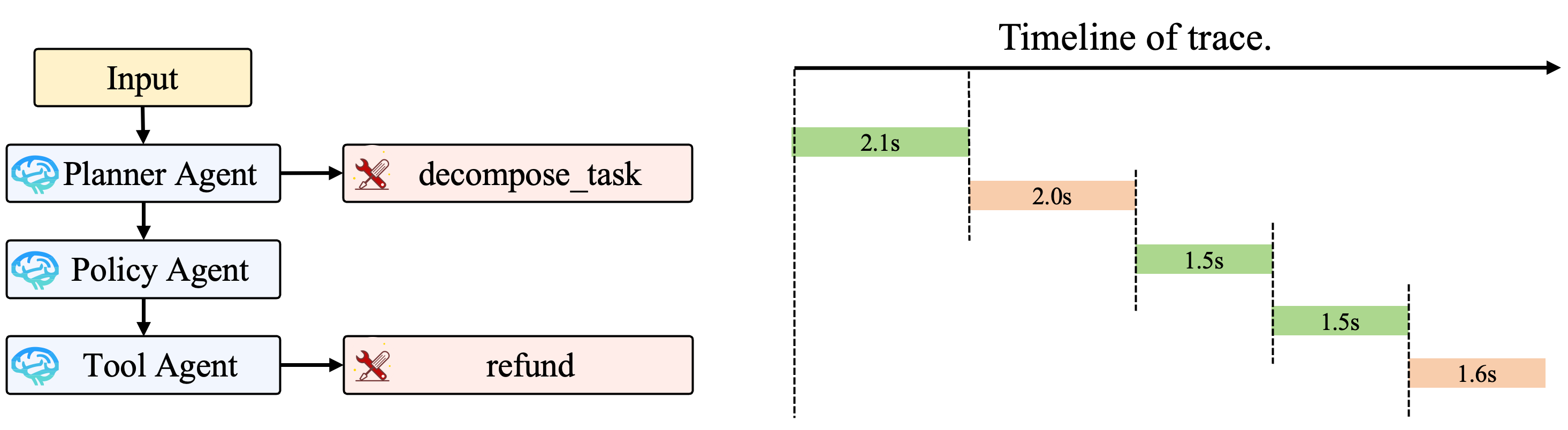}  
    \caption{\redd{Trace data of the agent system in Fig. \ref{fig:agentops-example}.}}
    \label{fig:trace}  
\end{figure}

In traditional monitoring data, metrics, logs, and traces collected using OpenTelemetry \cite{opentelemetry2019} exist within both microservice systems and agent systems.

\red{First, with respect to metrics, traditional microservice systems typically focus on system-level metrics and Application Performance Monitoring (APM) metrics. In contrast, agent systems that incorporate LLM-based agents introduce additional observability requirements, particularly due to the use of LLMs, external tools, and retrieval components. Based on current agent-system applications, we summarize four commonly used categories of metrics: system metrics, APM metrics, cost-related metrics, and RAG-related metrics. It should be noted that these categories are not exhaustive. Instead, they are intended to capture representative and frequently adopted metric types in practical agent-system deployments.}

\redd{Regarding traces, microservice traces typically capture interactions between services through API calls, whose interfaces and parameters are largely defined by fixed API schemas and application logic. In contrast, as illustrated by the customer refund scenario in Fig.~\ref{fig:agentops-example}, traces in agent systems need to capture not only the interaction paths among agents and tools, but also the dynamically generated inputs and outputs at each execution step. Fig.~\ref{fig:trace} further illustrates the trace data collected from this scenario, including the refund request, the intermediate output produced by the Policy Agent, the subsequent API call issued by the Tool Agent, and the verification result generated by the Verifier Agent. Since such intermediate contents are often generated or interpreted by LLMs, they may vary across executions and introduce additional uncertainty into the execution process. Therefore, these step-level inputs and outputs, together with agent-to-agent and agent-to-tool interactions, constitute essential trace data in agent systems and provide critical context for understanding and diagnosing their execution behavior.}

\begin{figure}[t]
    \centering
    \includegraphics[width=0.49\textwidth]{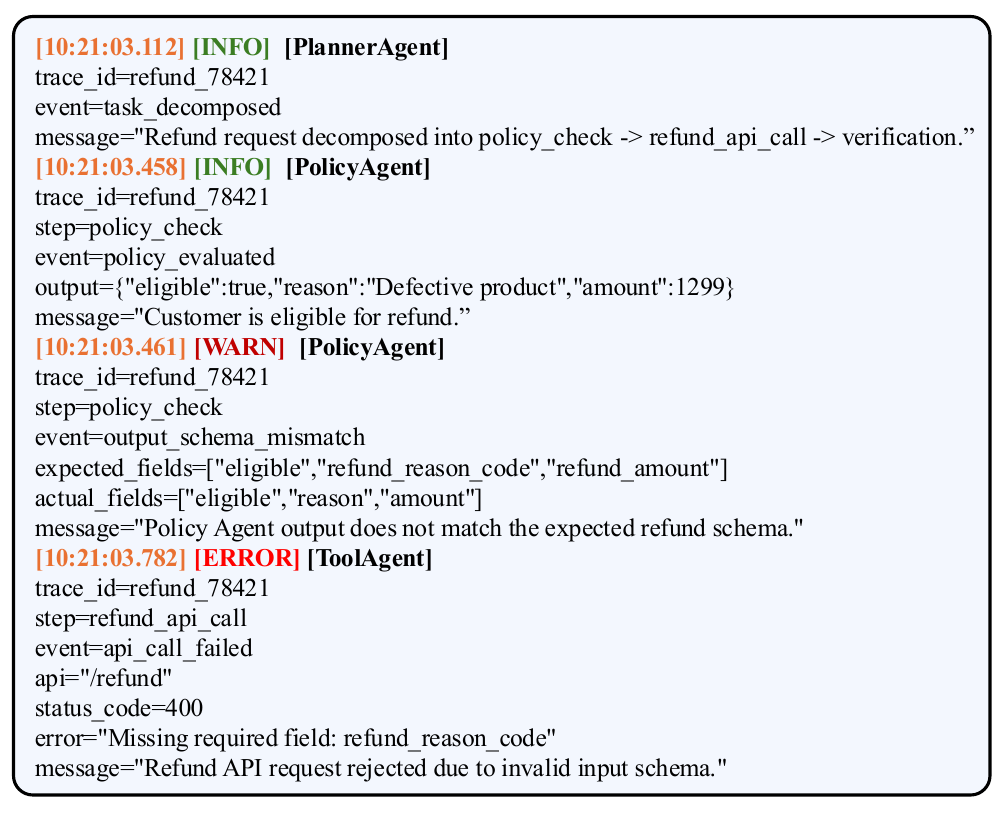}  
    \caption{\redd{Log data of the agent system in Fig. \ref{fig:agentops-example}.}}
    \label{fig:log}  
\end{figure}

When it comes to logs, as shown in Fig.~\ref{fig:log}, microservice systems and agent systems are similar. The logs in a microservice system record the overall behavior of services, whereas in an agent system, they capture the behavior of agents.



\begin{figure}[t]
    \centering
    \includegraphics[width=0.4\textwidth]{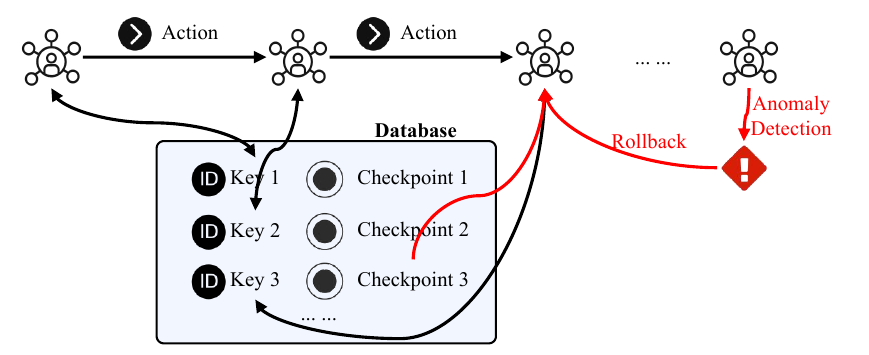}  
    \caption{Collection of checkpoint data and its use in rollback processes.}
    \label{fig:checkpoint}  
\end{figure}



\subsection{\redd{Monitoring Tools}}

\begin{table}[t]
\centering
\caption{\red{Comparison of representative observability tools. The listed tools are representative examples, and many unlisted tools provide similar functionalities.}}
\label{tab:observe}
\resizebox{0.48\textwidth}{!}{
\begin{tabular}{ccccccc} 
\toprule
\multicolumn{1}{l}{}   & \multicolumn{4}{c}{\textbf{Metric}}                                                                                & \multirow{2}{*}{\textbf{Log}} & \multirow{2}{*}{\textbf{Trace}}  \\ 
\cmidrule{2-5}
\textbf{Tools}         & \textbf{System}     & \textbf{Cost}       & \textbf{RAG}        & \textbf{Performance} &                               &                                  \\ 
\midrule
\textbf{LangDB \cite{langdb2025}}        & \ding{51} & \ding{51} & \ding{55} & \ding{55}  & \ding{51}    & \ding{51}       \\
\textbf{LangFuse \cite{langfuse2025}}      & \ding{51} & \ding{51} & \ding{51} & \ding{51}  & \ding{51}    & \ding{51}       \\
\textbf{MLFlow \cite{mlflow2025}}        & \ding{55} & \ding{51} & \ding{55} & \ding{51}  & \ding{51}    & \ding{51}       \\
\textbf{Helicone \cite{helicone2025}}      & \ding{51} & \ding{51} & \ding{55} & \ding{51}  & \ding{51}    & \ding{51}       \\
\textbf{LangWatch \cite{langwatch2025}}     & \ding{51} & \ding{51} & \ding{55} & \ding{51}  & \ding{51}    & \ding{51}       \\
\textbf{LlamaTrace \cite{arize_llamatrace2025}}    & \ding{51} & \ding{51} & \ding{55} & \ding{51}  & \ding{51}    & \ding{51}       \\
\textbf{OpenLLMetry \cite{openllmetry2025}}   & \ding{51} & \ding{55} & \ding{55} & \ding{51}  & \ding{51}    & \ding{51}       \\
\textbf{Arize Phoenix \cite{arize_phoenix2025}} & \ding{51} & \ding{51} & \ding{55} & \ding{51}  & \ding{51}    & \ding{51}       \\
\textbf{Literal AI \cite{literalai2025}}    & \ding{51} & \ding{51} & \ding{55} & \ding{51}  & \ding{51}    & \ding{51}       \\
\textbf{Opik \cite{comet_opik2025}}          & \ding{51} & \ding{51} & \ding{55} & \ding{51}  & \ding{51}    & \ding{51}       \\
\textbf{OpenInference \cite{arize_openinference2025}} & \ding{51} & \ding{55} & \ding{55} & \ding{51}  & \ding{51}    & \ding{51}       \\
\textbf{TruLens \cite{trulens2025}}       & \ding{51} & \ding{51} & \ding{55} & \ding{51}  & \ding{51}    & \ding{51}       \\
\textbf{HoneyHive \cite{honeyhive2025}}     & \ding{51} & \ding{51} & \ding{55} & \ding{51}  & \ding{51}    & \ding{51}       \\
\textbf{PromptLayer \cite{promptlayer2025}}   & \ding{51} & \ding{51} & \ding{55} & \ding{51}  & \ding{51}    & \ding{51}       \\
\textbf{\red{AgentOpsMonitor} \cite{agentops}}      & \ding{51} & \ding{51} & \ding{55} & \ding{51}  & \ding{51}    & \ding{51}       \\
\textbf{DeepEval \cite{DeepEval}}      & \ding{51} & \ding{51} & \ding{55} & \ding{51}  & \ding{51}    & \ding{51}       \\
\textbf{\red{LangSmith} \cite{langchain_langsmith_platform}}      & \ding{51} & \ding{51} & \ding{51} & \ding{51}  & \ding{51}    & \ding{51}       \\
\bottomrule
\end{tabular}
}
\end{table}

Observability tools for LLM-based agent systems are rapidly evolving. Most of these tools follow the general observability paradigm of collecting metrics, logs, and traces, while further extending it with agent-specific functions such as dataset management, experiment tracking, evaluations, prompt optimization, and deployment management. Table~\ref{tab:observe} compares the functionalities of representative tools. LangDB~\cite{langdb2025}, the first observability tool implemented entirely in Rust, emphasizes runtime efficiency and integrates router optimization for cost control. LangFuse~\cite{langfuse2025} supports OpenTelemetry integration and is among the most active open-source tools in this area. Helicone~\cite{helicone2025} combines observability with cache management to reduce latency and resource consumption, and improves reliability through mechanisms such as gateway fallback. HoneyHive~\cite{honeyhive2025} supports distributed tracing and multimodal systems, allowing users to define custom spans for fine-grained analysis. PromptLayer~\cite{promptlayer2025}, originally designed for prompt optimization, incorporates observability features such as prompt ranking. TruLens~\cite{trulens2025}, provided as a Python package, integrates with frameworks such as LlamaIndex and supports iterative optimization with human feedback. OpenLLMetry~\cite{openllmetry2025} follows the OpenTelemetry standard and is compatible with multiple frameworks, but provides limited support for prompt optimization and evaluation. LangWatch~\cite{langwatch2025} and Literal AI~\cite{literalai2025} provide standard observability, evaluation, and development functionalities, with LangWatch also available as an MCP server. MLFlow~\cite{mlflow2025}, originating from traditional machine learning workflows, supports custom metrics for agent systems. DeepEval~\cite{DeepEval} primarily focuses on evaluation rather than observability, whereas \redd{AgentOpsMonitor}~\cite{agentops} emphasizes operational management in addition to observability. \redd{LangSmith~\cite{langchain_langsmith_platform} integrates tracing, evaluation, and dataset management, with particular strengths in debugging and optimizing LangChain-based agent workflows.}

\subsection{Future Directions}



\red{Current observability for agent systems is still largely designed for monitoring. 
However, effective observability should serve a broader purpose: it should support anomaly detection, root cause localization, and failure resolution throughout the lifecycle of agent execution. 
As discussed earlier in section \ref{sec:diff}, existing observable data in agent systems, such as metrics, logs, and traces, is often unreliable, incomplete, or semantically insufficient. 
As a result, these signals may not even provide adequate support for comprehensive anomaly detection, let alone precise localization and automated recovery.}

\red{Therefore, one promising direction is to enrich the types of observable data available to agent systems. 
In particular, model data and checkpoint data can serve as two important extensions to existing observability signals. 
As data security concerns continue to grow, an increasing number of online agent systems are adopting locally deployed open-source LLMs as inference engines~\cite{datasecurity}. 
This deployment paradigm enables agent systems to move beyond black-box observation, where only inputs and outputs are available, and instead collect white-box model data from internal states, such as hidden-layer representations and token logits~\cite{huang2024opera,SAPLMA}. 
Compared with input-output records alone, such model-level data can provide more fine-grained evidence for detecting abnormal behaviors that originate within the LLM itself, thereby improving the observability of agent reasoning and generation processes.}


\redd{Checkpoint data provides another complementary type of observable data. Compared with traditional microservice systems, agent systems often provide greater controllability, since their execution state at a given moment can be reproduced from recorded data. Therefore, checkpoints can periodically capture the execution state of an agent system, including memory, environment, intermediate results, and other runtime information. As shown in Fig.~\ref{fig:checkpoint}, once an anomaly is detected, the system can roll back to a previously safe checkpoint, incorporate feedback, and resume execution from a corrected state. In coding-agent systems, repository revisions provide a concrete realization of such checkpoint-aware observability. For example, ARC~\cite{arc} uses Git revision history to record intermediate repository states, agent modifications, and associated test outcomes during test-driven software construction, providing an auditable execution history for subsequent diagnosis and recovery.}

\red{In addition to expanding the categories of observable data, future observability should also improve the quality of such data. 
Rather than relying solely on raw metrics, logs, traces, model data, or checkpoint records, agent systems require processed and risk-aware signals that more directly reflect potential failures. 
For example, hallucination probability inferred from execution traces or token-level uncertainty estimated from logits can provide more actionable evidence than raw records alone. 
Such high-quality observability signals can help agent frameworks detect, localize, and recover from failures more efficiently, enabling agent systems to remain stable, reliable, and continuously improvable.} \redd{Another promising direction is to transform raw observations into specification-grounded semantic signals. For example, WebTestPilot~\cite{teoh2026webtestpilot} represents GUI states with symbolic variables and infers step-wise preconditions and postconditions from natural-language specifications, providing executable semantic oracles for validating multi-step executions.}

\section{How to Detect Anomalies in Agent Systems?}
\label{sec:ad}


After obtaining sufficiently diverse and comprehensive monitoring data through multi-dimensional observation, these high-quality data can be utilized to perform operations on the agent system. Among these operations, anomaly detection plays a crucial role, as promptly identifying anomalies within the system is essential for effectively guiding subsequent root cause localization and resolution. \redd{As illustrated in Fig. \ref{fig:agentops-example}, anomaly detection transforms the monitored degradation and abnormal execution traces into actionable signals by identifying the Policy Agent, Tool Agent, and Verifier Agent, together with their corresponding abnormal steps.} This narrows the scope of subsequent diagnosis before the underlying cause is further investigated. This section primarily focuses on the methods of anomaly detection. A summary of anomaly detection and mitigation in agent systems is presented in Table~\ref{tab:anomaly-detection-prevention}. Below, we primarily introduce detection methods for some common anomalies.

\begin{table*}[t]
\centering
\caption{A summary of anomaly detection and mitigation methods for agent systems. Here, mitigation denotes targeted mechanisms or auxiliary models designed to address specific anomaly categories, whereas Section~\ref{sec:resolution} discusses anomaly resolution from the perspective of the overall AgentOps framework.}
\label{tab:anomaly-detection-prevention}
\resizebox{\textwidth}{!}{
\begin{tabular}{cccccccc}
\toprule
\textbf{Anomaly Level} 
& \textbf{Anomaly Type} 
& \textbf{Method Category} 
& \textbf{Method} 
& \textbf{Input} 
& \textbf{Output} 
& \textbf{Detection} 
& \textbf{Mitigation} \\
\midrule

\multirow{16}{*}{\textbf{Intra-Agent}}
& \multirow{7}{*}{\textbf{Reasoning Anomalies}}
& \multirow{3}{*}{White-box}
& SAPLMA~\cite{SAPLMA} 
& LLM Parameters 
& Anomaly Probability 
& \ding{51} & \ding{55} \\
& & 
& OPERA~\cite{huang2024opera} 
& Attention Map 
& Penalty-regenerated Response 
& \ding{51} & \ding{51} \\
& & 
& Honesty~\cite{yang2024alignment} 
& LLM Parameters (Finetuning) 
& Revised Response 
& \ding{51} & \ding{51} \\
\cmidrule{3-8}

& 
& \multirow{2}{*}{Grey-box}
& LURE~\cite{zhou202LURE} 
& Token Logits, Revisor Model 
& Revised Response 
& \ding{51} & \ding{51} \\
& & 
& Conformal~\cite{quach2023conformal} 
& Token Logits 
& High-quality Response 
& \ding{51} & \ding{51} \\
\cmidrule{3-8}

& 
& \multirow{2}{*}{Black-box}
& Debate~\cite{debate} 
& Multiple Agents 
& Debated Response 
& \ding{51} & \ding{51} \\
& & 
& CoK~\cite{cok} 
& Multiple Data Sources 
& High-quality Response 
& \ding{51} & \ding{51} \\
\cmidrule{2-8}

& \multirow{2}{*}{\textbf{Action Anomalies}}
& \multirow{2}{*}{MCP}
& AI-Infra-Guard~\cite{ai-infra-guard} 
& AI Component 
& MCP Risks 
& \ding{51} & \ding{55} \\
& & 
& MCP Guardian~\cite{kumar2025mcpguardian} 
& AI Component 
& MCP Risks 
& \ding{51} & \ding{55} \\
\cmidrule{2-8}

& \multirow{4}{*}{\textbf{Memory Anomalies}}
& \multirow{2}{*}{Short-term Memory}
& PI~\cite{PI} 
& Long Context 
& Response 
& \ding{55} & \ding{51} \\
& & 
& CoA~\cite{CoA} 
& Long Context 
& Response 
& \ding{55} & \ding{51} \\
\cmidrule{3-8}

& 
& \multirow{2}{*}{Long-term Memory / RAG}
& ReDeep~\cite{sun2024redeep} 
& Attention, LLM Parameters 
& Hallucination Results 
& \ding{51} & \ding{51} \\
& & 
& LRP4RAG~\cite{hu2024lrp4rag} 
& Token Logits 
& Anomaly Probability 
& \ding{51} & \ding{55} \\
\cmidrule{2-8}

& \multirow{2}{*}{\textbf{Security Anomalies}}
& \multirow{2}{*}{Graph-based}
& GUARDIAN~\cite{zhou2025guardian} 
& Agent Graph 
& Anomalous Position 
& \ding{51} & \ding{55} \\
& & 
& SentinelAgent~\cite{he2025sentinelagent} 
& Agent Graph 
& Anomalous Position 
& \ding{51} & \ding{55} \\

\midrule

\multirow{14}{*}{\textbf{Inter-Agent}}
& \multirow{6}{*}{\textbf{Task Specification Anomalies}}
& \multirow{3}{*}{Classifier-based}
& SpecValidator~\cite{akli2026defective} 
& Task Description 
& Defect Type / Probability 
& \ding{51} & \ding{55} \\
& & 
& Ambig-SWE~\cite{vijayvargiya2026ambig} 
& Task Instruction 
& Underspecification Label / Clarification Questions 
& \ding{51} & \ding{51} \\
& & 
& CLAMBER~\cite{zhang2024clamber} 
& User Query 
& Ambiguity Label / Clarifying Question 
& \ding{51} & \ding{51} \\
\cmidrule{3-8}

& 
& \multirow{3}{*}{Uncertainty-based}
& Ask-or-Assume~\cite{edwards2026ask} 
& Task Instruction, Agent State 
& Ask-or-Execute Decision 
& \ding{51} & \ding{51} \\
& & 
& Semantic Entropy~\cite{kuhn2023semantic} 
& Sampled Interpretations 
& Semantic Uncertainty Score 
& \ding{51} & \ding{55} \\
& & 
& SelfCheckGPT~\cite{manakul2023selfcheckgpt} 
& Sampled Responses 
& Consistency Score 
& \ding{51} & \ding{55} \\
\cmidrule{2-8}

& \multirow{6}{*}{\textbf{Orchestration Anomalies}}
& \multirow{3}{*}{Single-step}
& Introspective~\cite{liang2024introspective} 
& Token Logits 
& High-quality Plan 
& \ding{51} & \ding{51} \\
& & 
& API-bank~\cite{li2023apibank} 
& LLM Parameters (Finetuning) 
& High-quality Plan 
& \ding{55} & \ding{51} \\
& & 
& ReAct~\cite{react} 
& Prompt 
& High-quality Plan 
& \ding{51} & \ding{51} \\
\cmidrule{3-8}

& 
& \multirow{3}{*}{Multi-step}
& ToolLLM~\cite{qin2023toolllm} 
& LLM Parameters (Finetuning) 
& High-quality Plan 
& \ding{55} & \ding{51} \\
& & 
& Reflexion~\cite{shinn2023reflexion} 
& External Feedback 
& High-quality Plan 
& \ding{51} & \ding{51} \\
& & 
& CodeAct~\cite{codeact} 
& Tool Description 
& Code 
& \ding{55} & \ding{51} \\
\cmidrule{2-8}

& \multirow{2}{*}{\textbf{Communication Anomalies}}
& \multirow{2}{*}{Redundancy}
& AgentPrune~\cite{zhang2024agentprune} 
& Agent Graph 
& Optimized Agent Graph 
& \ding{51} & \ding{51} \\
& & 
& G-Designer~\cite{zhang2024gdesigner} 
& Agent Graph 
& Optimized Agent Graph 
& \ding{51} & \ding{51} \\

\bottomrule
\end{tabular}
}
\vspace{0.5em}
\begin{minipage}{0.98\textwidth}
\footnotesize
\red{\textit{Note.} Termination anomalies are not included in this table because they are typically easy to observe or can be detected through rule-based methods, such as checking task success rates or the number of task execution steps.}
\end{minipage}
\end{table*}

\subsection{Detection Methods}

\subsubsection{Detecting Reasoning Anomalies}



The methods for detecting and mitigating reasoning anomalies can be categorized into white-box, grey-box, and black-box based on the type of input information used. White-box approaches utilize the token sequence, the associated probabilities of each token, and the model parameters. In contrast, grey-box methods do not use model parameters. Lastly, black-box techniques rely solely on the token sequence.

\begin{itemize}[leftmargin=*]
    \item \textbf{White-box}. SAPLMA \cite{SAPLMA} posits that LLM parameters contain implicit clues of hallucinations and proposes a parameter-based classifier, trained on a large and diverse dataset, to detect hallucinations in a zero-shot manner. In contrast, OPERA \cite{huang2024opera} addresses hallucinations during the inference phase by identifying the “partial over-trust” phenomenon in self-attention maps, mitigating it through a column-wise metric, probabilistic penalties, and a rollback strategy. Meanwhile, Honesty \cite{yang2024alignment} attributes hallucinations to queries beyond the model's knowledge boundaries, advocating for a model that is honest about its limitations by introducing evolutionary metrics and an IDK (“I don’t know”) response, alongside techniques like prompt engineering and supervised fine-tuning to resolve such anomalies.
    \item \textbf{Grey-box}. LURE \cite{zhou202LURE} analyzes three primary causes of hallucinations—co-occurrence, uncertainty, and object position—and uses LLMs to generate samples for training a lightweight revisor that can directly detect and correct responses during the inference phase. Conformal \cite{quach2023conformal}, inspired by conformal prediction, calculates the quality of sampled responses based on token sequence logits and discards them according to a stopping rule.
    \item \textbf{Black-box}. Debate \cite{debate} reduces hallucinations by having multiple model instances iteratively evaluate and update their answers to the same question, drawing on the perspective of The Society of Mind. CoK \cite{cok} enhances factual accuracy by enabling the LLM to actively query and integrate information from multiple external knowledge sources during inference, iteratively reviewing and correcting the reasoning chain.
\end{itemize}

\subsubsection{Detecting Action Anomalies}

\redd{Action anomalies are primarily categorized into function call anomalies and Model Context Protocol (MCP) anomalies. Function call anomalies are relatively straightforward to address due to their clear error messages and direct handling, so they are not elaborated here. In contrast, for more complex MCP anomalies, AI-Infra-Guard \cite{ai-infra-guard} employs agent-based techniques to detect potential risks within MCP, such as tool poisoning attacks. The core mechanism of MCP Guardian \cite{kumar2025mcpguardian} involves embedding a ``security-first'' middleware layer within MCP communication, which provides four key functions: authentication, rate limiting, detailed logging of call information, and deep inspection of packet contents, thereby enhancing the communication between the client (MCP client) and the data source/tool server (MCP server).}

\subsubsection{Detecting Memory Anomalies}


Since an agent's memory is divided into short-term memory and long-term memory, memory anomalies can be categorized into short-term memory anomalies and long-term memory anomalies.

\begin{itemize}[leftmargin=*]
    \item \textbf{Short-Term Memory}. To address the challenge of extending the context window length, PI \cite{PI} proposes position interpolation, which scales the entire window through position encoding. In contrast, CoA \cite{CoA} adopts a multi-agent collaborative framework, where different agents handle separate chunks and a managing agent integrates the results, mitigating the ``lost in the middle'' issue commonly observed in traditional fine-tuning methods.
    \item \textbf{Long-Term Memory}. ReDeep \cite{sun2024redeep} mitigates hallucinations by computing two scores—an external context score and a parametric knowledge score—to modulate the contributions of Knowledge FFNs and Copying Heads within the residual stream. LRP4RAG \cite{hu2024lrp4rag}, on the other hand, employs the Layer-wise Relevance Propagation (LRP) algorithm, performing LRP backward after the generator outputs logits to obtain a relevance matrix, which is then fed into a pre-trained classifier to determine hallucinations.
\end{itemize}

\subsubsection{Detecting Security Anomalies}

Security anomalies pose significant threats to system security, and detection methods often rely on graph-based approaches. GUARDIAN \cite{zhou2025guardian} highlights that traditional multi-agent voting approaches overlook dependencies between agents. To address this, it models the agent system as a graph, using an encoder-decoder architecture to compress and reconstruct the entire graph, with anomalies detected based on the reconstruction score. SentinelAgent \cite{he2025sentinelagent}, on the other hand, tackles multi-agent coordination risks by proposing a three-layer hierarchical detection method that operates from global to detailed levels.








\subsubsection{Detecting Task Specification Anomalies}





\redd{Task specification anomalies arise when user instructions are ambiguous, incomplete, or underspecified, making it difficult for an agent to correctly infer the intended goals, constraints, or execution requirements. Unlike most execution-time anomalies, such failures may already be present before the agent begins acting and can subsequently propagate into planning and execution. Existing detection methods can be broadly categorized into \emph{classifier-based} and \emph{uncertainty-based} approaches.}

\begin{itemize}[leftmargin=*]
    \item \redd{\textbf{Classifier-based.} Classifier-based methods directly assess whether a task description contains specification defects. SpecValidator~\cite{akli2026defective} detects defective task descriptions and predicts the corresponding defect type or probability. Ambig-SWE~\cite{vijayvargiya2026ambig} focuses on underspecified instructions in software-engineering tasks; in addition to identifying underspecification, it generates clarification questions to acquire missing information before execution. Similarly, CLAMBER~\cite{zhang2024clamber} studies ambiguous information needs in user queries and enables models to both recognize ambiguity and request clarification. These methods explicitly model task-specification quality and are therefore suitable for intercepting malformed or incomplete requests before they propagate to downstream agent behaviors.}

    \item \redd{\textbf{Uncertainty-based.} Uncertainty-based methods, in contrast, infer specification anomalies from the uncertainty or inconsistency induced by an instruction. Ask-or-Assume~\cite{edwards2026ask} jointly considers the task instruction and the current agent state to determine whether the agent should proceed with execution or first ask for clarification. More general uncertainty-estimation methods can also provide useful signals: Semantic Entropy~\cite{kuhn2023semantic} measures semantic uncertainty across sampled interpretations, while SelfCheckGPT~\cite{manakul2023selfcheckgpt} evaluates consistency among sampled responses. Such methods are particularly useful when ambiguity cannot be captured by predefined defect categories, as an underspecified task may admit multiple plausible interpretations and consequently produce high uncertainty or inconsistent agent behaviors.}
\end{itemize}

\subsubsection{Detecting Orchestration Anomalies}

Orchestration anomalies can be categorized based on their occurrence location into single-step and multiple-step anomalies.

\begin{itemize}[leftmargin=*]
    \item \textbf{Single-step}. Introspective \cite{liang2024introspective} combines RAG recall with human feedback-based Conformal Prediction to address common-sense reasoning issues in LLMs. API-bank \cite{li2023apibank} improves LLM planning capabilities by finetuning the model with dialogue data collected from API calls, using a data-driven approach. ReAct \cite{react}, on the other hand, introduces a reasoning and acting framework that splits planning into two stages—reasoning and acting—and implements it through prompt engineering, providing the model with opportunities for reflective error correction.
    \item \textbf{Multiple-step}. ToolLLM \cite{qin2023toolllm} generates both single-tool and multi-tool instructions based on APIs and further proposes a novel depth-first search–based decision tree to construct coherent planning paths. Reflexion \cite{shinn2023reflexion} leverages external feedback—such as simple binary environment feedback, predefined heuristics for common failure cases, and self-evaluation mechanisms like binary classification using LLMs (for decision-making) or self-written unit tests (for programming)—to enable agents to reflect on past execution steps and optimize subsequent trajectories. CodeAct, in turn, formulates the entire planning process as executable Python code, fundamentally reducing hallucinations in tool usage by ensuring precise parameter passing.

\end{itemize}

\subsubsection{Detecting Communication Anomalies}

\redd{Communication anomalies primarily arise from redundant messages caused by the scaling of agent systems. To address this issue, AgentPrune \cite{zhang2024agentprune} models the system as a graph and trains a low-rank-principle-guided graph mask to focus on key communication areas, thereby mitigating communication redundancy and improving efficiency. G-Designer \cite{zhang2024gdesigner}, from a more macro perspective, argues that the optimal communication topology varies depending on the task. To this end, a graph auto-encoder compresses the entire agent system and reconstructs an optimal topology.}


\redd{The primary cause of communication anomalies is the continuous expansion of agent systems, leading to an increasing volume of transmitted messages. However, many of these messages are unnecessary, and excessive redundancy can result in task failure.}

\subsection{Future Directions}

\red{Although existing studies have proposed various detection methods for reasoning, action, memory, orchestration, and communication anomalies, most of them remain category-specific, relying on distinct input signals, detection granularities, and modeling assumptions. Therefore, an important future direction is to move from anomaly-specific detectors toward unified and generalizable detection frameworks. Such frameworks should integrate model-internal states, reasoning traces, tool-call records, memory access patterns, and agent interaction structures, enabling different anomaly patterns to be characterized within a shared representation space. This would improve the applicability of anomaly detection methods across diverse tasks, architectures, and deployment environments.}

\red{Moreover, future research should further bridge the intra-agent and inter-agent levels. Existing methods typically focus either on anomalies within individual agents or on anomalies emerging from multi-agent collaboration. However, in real-world agent systems, reasoning, memory, or action anomalies at the individual-agent level may further manifest as orchestration failures or communication anomalies at the system level. Cross-level detection methods should therefore jointly model the internal states of individual agents and the interaction structures among multiple agents, so as to identify local abnormal signals at an early stage and capture their system-level manifestations.}

\red{Finally, online anomaly detection for long-horizon tasks represents another key direction. Practical agent systems often operate continuously in dynamic environments, where anomalies may gradually accumulate through iterative planning, tool use, memory updates, and communication processes. Compared with post-hoc detection based on final outputs or offline logs, online detection enables continuous monitoring of agent behaviors during execution, allowing potential abnormal trends to be identified in a timely manner and providing more fine-grained signals for subsequent root-cause analysis and anomaly resolution.}

\section{How to Locate the Root Cause of Agent System Anomalies?}

\subsection{Definition of Root Cause Localization in AgentOps}


\red{In the context of AgentOps, root cause localization---also referred to as failure attribution---aims to identify which component causes an anomaly and at which point in the execution it first occurs, by analyzing the execution trajectory of an agent system. \redd{Continuing the example in Fig. \ref{fig:agentops-example}, anomaly detection flags multiple abnormal agents and steps, whereas root cause localization further distinguishes the upstream cause from its downstream manifestations: the schema change in the Policy Agent is identified as the root cause of the subsequent Tool Agent API failure and Verifier Agent failure.} In this work, root cause localization primarily focuses on anomalies that arise during the execution phase. }



\subsection{Root Cause Localization Methods}

\begin{table}[t]
\centering
\footnotesize
\caption{\red{Summary of root cause localization methods in agent systems.}}
\label{tab:failure-attribution-summary}
\setlength{\tabcolsep}{3pt}
\renewcommand{\arraystretch}{1.0}
\begin{tabularx}{\columnwidth}{
>{\raggedright\arraybackslash}p{1.8cm}
>{\raggedright\arraybackslash}X
}
\toprule
\textbf{Category} & \textbf{Attribution Basis and Methods} \\
\midrule

\textbf{Replay \& Spectrum}
& Success/failure trajectories; state--action--agent tuples; statistical suspiciousness.
\newline
\emph{Methods:} FAMAS~\cite{ge2025introducing} \\

\midrule

\textbf{Structured Trace}
& Structured traces, communication logs, and causal/dependency graphs; information-flow and propagation analysis.
\newline
\emph{Methods:} GraphTracer~\cite{zhang2025graphtracer}; AgenTracer~\cite{agentracer} \\

\midrule

\textbf{LLM Attribution}
& Natural-language logs, segmented trajectories, and failure taxonomies; semantic judgment and taxonomy-guided reasoning.
\newline
\emph{Methods:} Who\&When~\cite{icml2025}; AgentFail~\cite{agentfail} \\

\bottomrule
\end{tabularx}
\vspace{0.2em}

\end{table}

\subsubsection{Trajectory Replay \& Spectrum Analysis}

This category quantifies the “suspiciousness” of actions or agents by comparing the differences between successful and failed trajectories. FAMAS \cite{ge2025introducing} introduces a spectral-analysis framework based on trajectory abstraction, in which execution logs are decomposed into state–action–agent tuples. It further defines a hybrid scoring function that jointly considers failure frequency, internal repetitiveness, and action centrality. By ranking suspiciousness scores, the system can automatically identify the agents and key actions most likely responsible for the failure.



\subsubsection{\redd{Structured Trace Modeling}}

\redd{As system complexity increases, researchers have proposed structured trajectory modeling methods to capture long-term and cross-agent dependencies. GraphTracer \cite{zhang2025graphtracer} constructs an information dependency graph to explicitly model inter-agent information flow and identifies root nodes causing downstream errors via backward tracing. AgenTracer \cite{agentracer} generates synthetic failure trajectories through counterfactual replay and fault injection, training a small dedicated LLM to learn the who and when localization tasks. Ma et al. \cite{ma2025automatic} construct structured causal graphs and apply counterfactual reasoning to attribute system-level failures to specific agents and actions. SentinelAgent \cite{he2025sentinelagent} models inter-agent interactions and execution dynamics as graphs to capture anomalous coordination and behavioral patterns. More recently, StepFinder \cite{zhu2026stepfinder} models execution traces as temporal semantic sequences. It uses LLMs only for semantic feature construction and subsequently applies temporal modeling and attention mechanisms to capture sequential evolution and cross-step dependencies, enabling efficient step-level root cause localization.}

\subsubsection{LLM-based Attribution}

Another mainstream direction is to leverage large language models to perform “log adjudication” (LLM-as-a-judge). Zhang et al. \cite{icml2025} propose three inference paradigms: \textbf{All-at-once}, where the full failure log is provided and the LLM outputs the faulty agent and step in a single pass; \textbf{Step-by-step}, where log segments are fed incrementally and the LLM decides whether the current step is erroneous, terminating immediately upon detection; and \textbf{Binary Search}, which recursively bisects the log interval and uses iterative LLM judgments to progressively narrow down the error span. These strategies exploit LLMs’ semantic reasoning ability, enabling soft, inference-driven localization even in the absence of highly structured logs. 
Although experiments show that pure LLM-based judgment still falls short in step-level accuracy, its advantage lies in supporting semi-automated diagnostic workflows when combined with human oversight, providing interpretable, language-based insights that facilitate system debugging.

AgentFail \cite{agentfail} further observes that LLMs struggle to reason effectively over complex trajectories. To address the difficulty LLMs face in accurately diagnosing the root causes of complex failures, AgentFail proposes a fine-grained taxonomy of agent system failures, which is used to guide the structured knowledge representation of LLMs. In the AgentFail benchmark, incorporating this taxonomy into LLM guidance improves localization identification accuracy by approximately 15\%. AgentDebug \cite{agentdebug} similarly argues that failures in agent systems are closely tied to failure taxonomies. By analyzing a large number of cases, it identifies that different types of failures exhibit distinct tendencies across different stages of an agent system’s execution. Accordingly, AgentDebug integrates taxonomy-aware analysis at each step of the agent system to jointly assess system behavior. Upon detecting a failure, the agent replans its trajectory, enabling timely resolution.

\subsection{Future Directions}

\begin{figure}[t]
    \centering
    \includegraphics[width=0.47\textwidth]{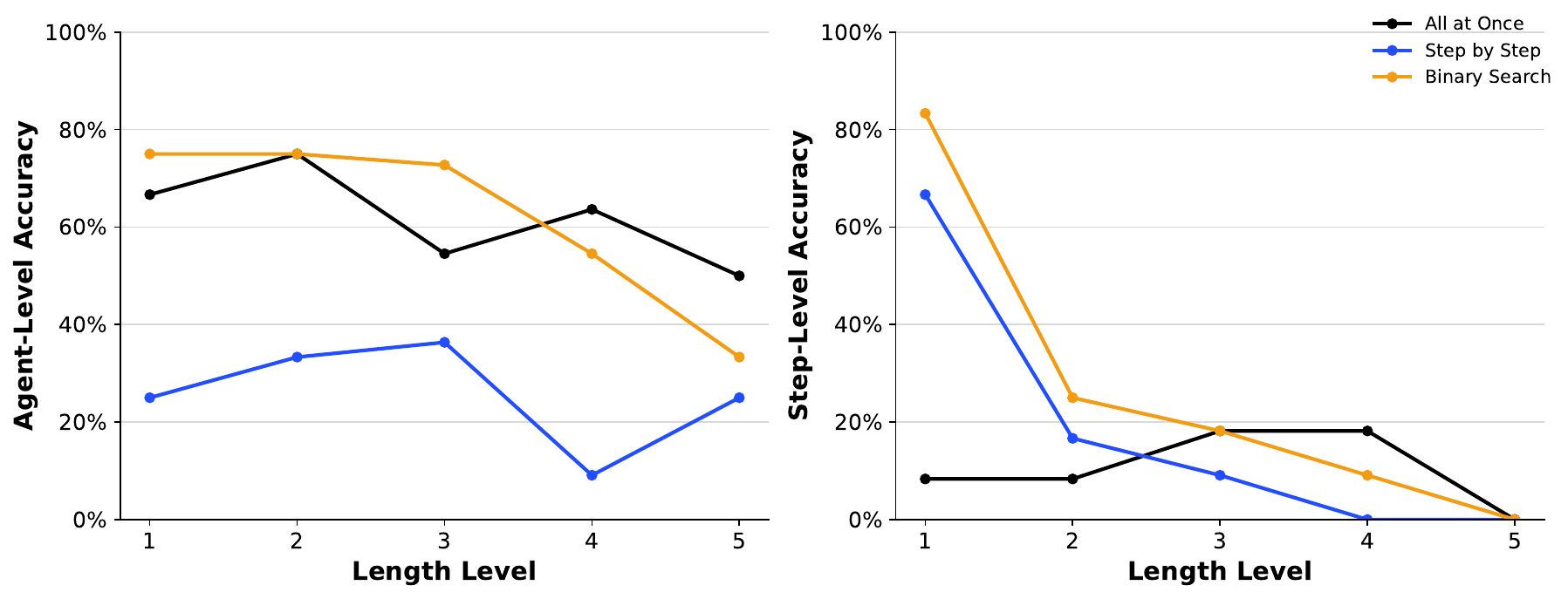}
    \caption{\redd{Root cause localization accuracy of different methods on the Who\&When dataset under varying context lengths (numbers of steps). Left: agent-level accuracy for identifying the erroneous agent; right: step-level accuracy for locating the erroneous step. All-at-once feeds the full trace to the LLM at once, Step-by-step processes one step at a time, and Binary-search locates the erroneous step via binary search over the full trace.}}
    \label{fig:diffcontextlength}
\end{figure}

\redd{Existing root cause localization methods can be broadly categorized into two groups: LLM-based methods and non-LLM-based methods, including statistical analysis and trace-based models. These two categories exhibit complementary strengths. LLM-based methods leverage the strong semantic reasoning capabilities of LLMs, enabling fine-grained fault localization at both the agent and step levels. However, their effectiveness degrades as the execution trace becomes longer. As shown in Fig.~\ref{fig:diffcontextlength}, where context length is measured by the number of steps in a trace, LLM-based methods become increasingly susceptible to the ``lost-in-the-middle'' problem, which can lead to hallucinated or incorrect attributions in long and complex agent trajectories. In contrast, non-LLM-based methods can learn statistical patterns and structural dependencies among agents and execution steps from historical traces, allowing them to efficiently identify suspicious regions even in long trajectories. Nevertheless, such methods typically provide coarse-grained predictions and lack the semantic reasoning capability required for precise root-cause localization, particularly at the step level.}

\redd{A promising future direction is to integrate two types of methods in a two-stage framework. The first stage uses a non-LLM-based model to narrow down suspicious regions, while the second stage applies an LLM-as-a-Judge to these regions for fine-grained attribution.}

\section{How to Resolve the Anomalies?}
\label{sec:resolution}


Once former steps have accurately pinpointed the root location, traditional operations can typically perform rapid recovery actions such as node switching or hardware maintenance to prevent further economic losses. However, the resolution process in agent systems differs substantially, both in its paradigms and methodologies. The following sections provide a detailed discussion of these distinctions.

\begin{table}[t]
\centering
\small
\caption{Targeted anomaly types of resolution methods. 
$\checkmark$ denotes a target, and -- denotes an indirect target.}
\label{tab:resolution-anomaly-targets}
\setlength{\tabcolsep}{1.8pt}
\renewcommand{\arraystretch}{0.95}
\resizebox{0.5\textwidth}{!}{
\begin{tabular}{lcccccccc}
\toprule
\textbf{Method} 
& \multicolumn{4}{c}{\textbf{Intra-agent}} 
& \multicolumn{4}{c}{\textbf{Inter-agent}} \\
\cmidrule(lr){2-5} \cmidrule(lr){6-9}
& \textbf{Reas.} 
& \textbf{Act.} 
& \textbf{Mem.} 
& \textbf{Sec.} 
& \textbf{Task} 
& \textbf{Comm.} 
& \textbf{Term.} 
& \textbf{Orch.} \\
\midrule

\multicolumn{9}{l}{\textit{Pre-execution preventive}} \\
Task Re-spec. 
& $\checkmark$ & $\checkmark$ & -- & -- & $\checkmark$ & -- & $\checkmark$ & $\checkmark$ \\

Plan Verif. 
& $\checkmark$ & $\checkmark$ & -- & -- & -- & -- & -- & $\checkmark$ \\

Action Restrict. 
& -- & $\checkmark$ & -- & $\checkmark$ & -- & -- & -- & -- \\


\midrule
\multicolumn{9}{l}{\textit{In-execution corrective}} \\
Obs.-grounded Replan. 
& $\checkmark$ & $\checkmark$ & -- & -- & -- & -- & $\checkmark$ & $\checkmark$ \\

Self-correction 
& $\checkmark$ & $\checkmark$ & -- & -- & -- & -- & -- & -- \\

Redundant Exec. \& Select. 
& $\checkmark$ & $\checkmark$ & -- & -- & -- & -- & -- & $\checkmark$ \\

\midrule
\multicolumn{9}{l}{\textit{Post-execution recovery}} \\
Rollback \& Re-exec. 
& $\checkmark$ & $\checkmark$ & $\checkmark$ & $\checkmark$ & $\checkmark$ & $\checkmark$ & $\checkmark$ & $\checkmark$ \\

Memory \& Skill Update 
& $\checkmark$ & $\checkmark$ & $\checkmark$ & -- & -- & -- & $\checkmark$ & $\checkmark$ \\

Policy \& Prompt Rev. 
& $\checkmark$ & $\checkmark$ & -- & $\checkmark$ & $\checkmark$ & -- & $\checkmark$ & $\checkmark$ \\


\bottomrule
\end{tabular}
}
\vspace{0.2em}

\emph{Note.} 
Reas.=Reasoning, Act.=Action, Mem.=Memory, Sec.=Security, 
Task=Task Specification, Comm.=Communication, Term.=Termination, 
Orch.=Orchestration.
\end{table}

\subsection{Resolution Methods}


\red{To systematically resolve anomalies in complex agentic systems, we categorize resolution strategies according to the stage at which the intervention is applied: \textit{pre-execution preventive resolutions}, \textit{in-execution corrective resolutions}, and \textit{post-execution recovery resolutions}. Rather than focusing on how anomalies are diagnosed, this taxonomy emphasizes how an agent system can actively resolve or mitigate failures. These three classes are complementary and are often combined in practical agent frameworks~\cite{react,wu2024autogen,sweagent}.}

\subsubsection{Pre-execution Preventive Resolutions}

\red{Pre-execution preventive resolutions aim to avoid anomalies before the agent commits to external actions. These strategies are especially useful when actions are costly, unsafe, or irreversible.}

\begin{itemize}[leftmargin=*]
    \item \noindent\textbf{Task Re-specification}. 
    A common way to prevent agent failures is to rewrite or refine the task before execution. This includes clarifying ambiguous goals, decomposing complex objectives, adding explicit constraints, specifying success criteria, and converting open-ended requests into executable subtasks. Prompting methods such as Chain-of-Thought, ReAct, and Tree-of-Thoughts make intermediate reasoning or alternative plans explicit, thereby reducing the chance that the agent starts from an ill-formed plan~\cite{cot,react,tot}.

    \item \noindent\textbf{Plan Verification}. 
    Before execution, the agent can generate a candidate plan and check whether it is feasible, safe, and consistent with user constraints. Plan verification may involve rule-based validators, critic models, verifier agents, or human approval. This strategy prevents invalid action sequences before they are carried out. Multi-agent systems such as AutoGen support planner--critic--executor patterns, where one agent proposes actions and another reviews them before execution~\cite{wu2024autogen}.

    \item \noindent\textbf{Action Space Restriction}. 
    Agent behavior can be made more reliable by restricting what actions are available before execution begins. This includes limiting tool access, enforcing typed tool schemas, requiring valid parameters, using sandboxed environments, and defining permission boundaries. Tool-use systems such as Toolformer and agent frameworks such as SWE-agent show that the design of tool interfaces and action spaces strongly affects the reliability of agent execution~\cite{toolformer,sweagent}.

\end{itemize}

\subsubsection{In-execution Corrective Resolutions}

\red{In-execution corrective resolutions intervene while the agent is reasoning, calling tools, receiving observations, or updating its state. Their purpose is to correct errors before the execution trajectory becomes unrecoverable.}

\begin{itemize}[leftmargin=*]
    \item \noindent\textbf{Observation-grounded Replanning}. 
    During execution, new observations may contradict the original plan or reveal that the current trajectory is ineffective. The agent can resolve such deviations by revising its plan based on the latest environment feedback or tool outputs. ReAct is a representative framework that interleaves reasoning and acting, enabling the agent to update its next action after each observation~\cite{react}.

    \item \noindent\textbf{Self-correction}. 
    Self-correction allows the agent to identify and repair its own mistakes during execution. The agent may critique its intermediate reasoning, verify tool outputs, rerun failed steps, or generate revised actions. Methods such as Self-Refine and Reflexion demonstrate that iterative feedback, critique, and reflection can improve model behavior without updating model parameters~\cite{selfrefine,shinn2023reflexion}.

\item \noindent\redd{\textbf{Runtime Enforcement and Repair}. 
Runtime mechanisms can enforce structured constraints on tool calls, intermediate states, and model outputs to intercept unsafe or policy-violating actions before execution~\cite{agentspec}. They can also detect and repair incompatibilities between probabilistic LLM outputs and deterministic software components, such as malformed formats, invalid syntax, or interface mismatches~\cite{shao2026comfrey}, thereby preventing failures from propagating through the system.}

    \item \noindent\textbf{Redundant Execution and Selection}. 
    Instead of committing to a single trajectory, the system may generate multiple candidate actions or run multiple agents in parallel, then select the most reliable output through voting, ranking, or verifier scoring. This resolves stochastic reasoning errors by comparing independent hypotheses. Such mechanisms are commonly used in multi-agent debate and ensemble-style agent execution~\cite{selfconsistency,debate,wu2024autogen}.
\end{itemize}

\subsubsection{Post-execution Recovery Resolutions}

\red{Post-execution recovery resolutions are applied after an execution attempt has failed or completed with unsatisfactory performance. These strategies focus on restoring a valid state, repairing consequences, and reducing recurrence in future runs.}

\begin{itemize}[leftmargin=*]
    \item \noindent\textbf{Rollback and Re-execution}. 
    If an agent reaches an invalid or undesirable state, the system can restore a previous checkpoint (\redd{such as using Git revision history \cite{arc}}) and re-execute from that point with revised constraints or plans. This requires persistent state tracking, checkpointing, and replayable execution traces. Frameworks such as LangGraph support checkpointing and time-travel execution, which allow developers to resume or fork an agent run from earlier states~\cite{langgraph}.


    \item \noindent\textbf{Memory and Skill Update}. 
    After a failed or successful run, the agent can store useful experience as memory, corrected procedures, or reusable skills. This allows future executions to avoid similar failures. Reflexion stores verbal feedback from prior attempts, while Voyager builds a skill library that can be reused and composed in later tasks~\cite{shinn2023reflexion,voyager}.

    \item \noindent\textbf{Policy and Prompt Revision}. 
    When a failure reflects a recurring weakness, the system can revise prompts, policies, tool schemas, or orchestration rules after execution. This differs from runtime self-correction because the modification is persistent and affects future runs. Automatic prompt optimization methods and evolutionary prompt search provide systematic ways to update instructions based on task-level performance~\cite{prom11,prom12,prom13,evo1}.

\end{itemize}


\subsection{Future Directions}

\red{We identify three promising directions for advancing resolution mechanisms in agent systems. A key direction is to develop a unified framework that integrates preventive, in-process, and post-hoc resolution mechanisms. Rather than treating these stages as isolated modules, future agent systems should adaptively select suitable resolution strategies according to the task context, execution state, uncertainty, and risk of error propagation. Such a framework can better balance robustness, efficiency, and correction cost.}

\red{Another important direction is to improve agents' self-correction ability during execution. Existing methods such as Reflexion enable feedback-driven reflection, but they often rely on external feedback and may detect errors only after several subsequent steps. This delay can allow mistakes to propagate, especially in long-horizon or multi-agent settings. Future work should therefore enhance both rapid error detection, allowing agents to identify mistakes within only a few steps, and self-driven detection, enabling agents to monitor their own reasoning and actions without relying on external signals.}

\red{Post-hoc resolution is also crucial for enabling agents to learn from practice. In existing frameworks, such as Hermes-style systems, this process is often formulated as skill accumulation, where reusable capabilities are distilled from past trajectories. While recent work explores adversarial or self-learning mechanisms for skill discovery, most frameworks emphasize successful experiences and overlook failures. Future systems should distill not only successful skills but also reusable lessons from failed trajectories, such as constraints, diagnostic patterns, and failure-aware heuristics, so as to reduce repeated errors and improve long-term robustness.}


\section{Datasets and Benchmarks for AgentOps}
\label{Datasets}

\begin{table}[t]
\centering
\footnotesize
\caption{Comparison of different datasets and benchmarks. RC stands for root cause.}
\label{databenchmark}
\setlength{\tabcolsep}{2.5pt}
\renewcommand{\arraystretch}{1.05}
\begin{tabularx}{\columnwidth}{
>{\raggedright\arraybackslash}p{1.65cm}
>{\centering\arraybackslash}p{0.75cm}
>{\centering\arraybackslash}p{0.75cm}
>{\centering\arraybackslash}p{1.20cm}
>{\raggedright\arraybackslash}X
}
\toprule
\textbf{Dataset} 
& \textbf{\#Fail.} 
& \textbf{\#Types} 
& \textbf{Single RC}
& \textbf{Scalability and Agent Systems} \\
\midrule

Who\&When~\cite{icml2025}
& 184
& --
& $\checkmark$
& Partial. \emph{Systems:} 127 multi-agent LLM systems. \\

\midrule

MASFT~\cite{masft}
& 1642
& 14
& $\checkmark$
& Yes. \emph{Systems:} ChatDev, MetaGPT, HyperAgent, AppWorld, AG2, Magentic-One, OpenManus. \\

\midrule

TRAIL~\cite{trail}
& 841
& 21
& $\checkmark$
& No. \emph{Systems:} --. \\

\midrule

AgentFail~\cite{agentfail}
& 307
& 16
& $\checkmark$
& Partial. \emph{Systems:} Dify, Coze. \\

\midrule

Aegis~\cite{aegis}
& 24843
& 14
& $\checkmark$
& Yes. \emph{Systems:} LLM Debate, MacNet, AgentVerse, Dylan, SmolAgents, Magentic-One. \\

\midrule

Agent-Debug~\cite{agentdebug}
& 200
& 17
& $\checkmark$
& No. \emph{Systems:} ALFWorld, WebShop, GAIA. \\

\midrule

\redd{MP-Bench~\cite{mpbench}}
& 289
& --
& $\times$
& \redd{No. \emph{Systems:} MAgentic-One, CaptainAgent.} \\

\midrule

\redd{LongRCA Bench~\cite{longrca}}
& 1140
& --
& $\checkmark$
& \redd{No. \emph{Systems:} AutoGen-based and custom multi-agents.} \\

\bottomrule
\end{tabularx}
\end{table}

\begin{table}[t]
\centering
\small
\caption{\red{Coverage of existing studies under our proposed failure taxonomy. 
\textbf{\checkmark} denotes explicit coverage and -- denotes little or no substantial coverage.}}
\label{tab:taxonomy_mapping}
\setlength{\tabcolsep}{2.0pt}
\renewcommand{\arraystretch}{0.95}
\resizebox{\columnwidth}{!}{
\begin{tabular}{lcccc|cccc}
\toprule
\multirow{2}{*}{\textbf{Study}} 
& \multicolumn{4}{c|}{\textbf{Intra-agent}} 
& \multicolumn{4}{c}{\textbf{Inter-agent}} \\
\cmidrule(lr){2-5} \cmidrule(lr){6-9}
& \textbf{Reas.} 
& \textbf{Act.} 
& \textbf{Mem.} 
& \textbf{Sec.} 
& \textbf{Task} 
& \textbf{Comm.} 
& \textbf{Term.} 
& \textbf{Orch.} \\
\midrule

Who\&When~\cite{icml2025}      
& \checkmark & \checkmark & -- & -- & -- & -- & -- & -- \\

MASFT~\cite{masft}        
& \checkmark & -- & \checkmark & -- & \checkmark & \checkmark & \checkmark & -- \\

TRAIL~\cite{trail}       
& \checkmark & \checkmark & \checkmark & \checkmark & \checkmark & -- & \checkmark & \checkmark \\

AgentFail~\cite{agentfail}   
& \checkmark & \checkmark & \checkmark & -- & \checkmark & -- & \checkmark & \checkmark \\

Aegis~\cite{aegis}      
& \checkmark & -- & \checkmark & -- & \checkmark & \checkmark & \checkmark & -- \\

AgentDebug~\cite{agentdebug}    
& \checkmark & \checkmark & \checkmark & \checkmark & -- & -- & -- & -- \\

\redd{MP-Bench~\cite{mpbench}}    
& \checkmark & \checkmark &  -- &  -- & -- & -- & -- & -- \\

\redd{LongRCA~\cite{longrca}}    
& \checkmark & \checkmark &  -- &  -- & -- & -- & -- & \checkmark \\

\bottomrule
\end{tabular}
}
\vspace{0.2em}

\emph{Note.} 
Reas.=Reasoning, Act.=Action, Mem.=Memory, Sec.=Security, 
Task=Task Specification, Comm.=Communication, Term.=Termination, 
Orch.=Orchestration.
\end{table}

\begin{figure}[t]
    \centering
    \includegraphics[width=0.49\textwidth]{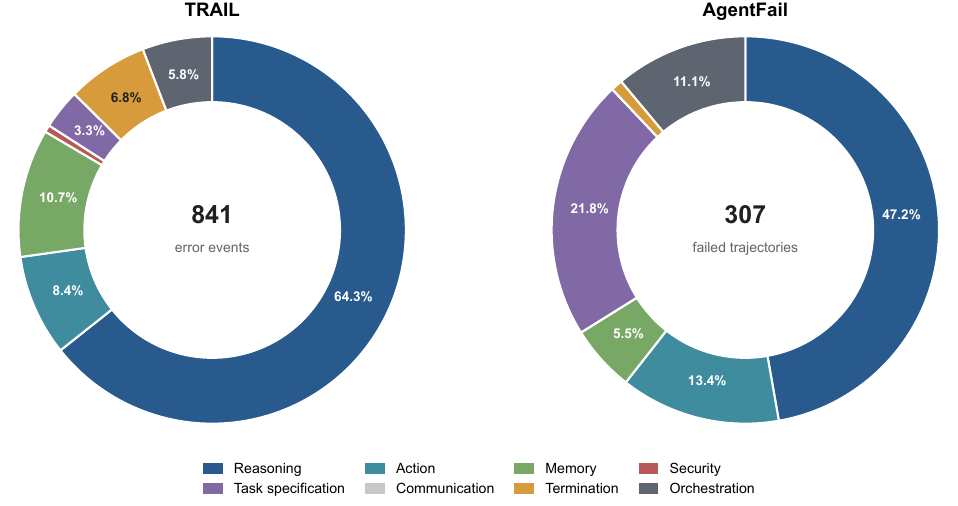}  
    \caption{\redd{Distribution of anomaly types in TRAIL and AgentFail under the AgentOps taxonomy.}}
    \label{fig:distribution}  
\end{figure}

\red{To better advance the development of AgentOps, recent efforts have introduced a growing number of datasets and benchmarks. Although AgentOps typically consists of four key stages, i.e., monitoring, anomaly detection, root cause localization, and resolution, not all stages are equally amenable to standardized dataset construction. In particular, monitoring and resolution are difficult to benchmark in a unified manner due to their strong dependence on practical deployment environments, system instrumentation, and downstream operational constraints. Meanwhile, as discussed earlier, anomaly detection is still largely characterized by methods tailored to specific and relatively narrow categories of anomalies, rather than a broadly standardized evaluation paradigm. Consequently, root cause localization, also referred to as failure attribution, has emerged as the most mature and standardized direction for dataset and benchmark development in AgentOps. Therefore, in this section, we primarily summarize existing datasets and benchmarks for root cause localization. As shown in Table~\ref{databenchmark}, most of these datasets are constructed by simulating trajectories using widely adopted agent systems in the community, such as GAIA~\cite{gaia}, ALFWorld~\cite{alfworld}, and WebShop~\cite{webshop}, followed by manual or automated annotation.} \redd{With the recent emergence of end-to-end testing frameworks for real-world Graphical User Interface (GUI) agents \cite{ahmed2026specops}, we expect trajectory-based datasets derived from realistic GUI interactions to continue expanding in both scale and diversity.}

Several clear trends can be observed. First, the number of failure instances has steadily increased. A larger number of failures enables richer failure combinations and allows for more fine-grained evaluation of different methods. Second, the types of failures have become increasingly diverse. Various agent failure taxonomies have been proposed, as comprehensive coverage of failure types is essential to assess whether an algorithm exhibits systematic bias toward specific error categories. \redd{However, despite the numerous taxonomies proposed in prior work (Table~\ref{tab:taxonomy_mapping}), existing efforts provide only limited coverage of the anomaly space. In particular, they predominantly focus on intra-agent anomalies, such as reasoning failures, while largely overlooking anomalies that arise from inter-agent interactions. As illustrated in Fig.~\ref{fig:distribution}, reasoning anomalies account for 47.2\%–64.3\% in the two analyzed datasets, indicating a strong bias toward intra-agent issues. Meanwhile, with the emergence of automated multi-agent system construction frameworks such as Hermes Agent, effective inter-agent collaboration is becoming increasingly critical, calling for a more comprehensive characterization of anomalies at the inter-agent level.} \redd{Third, recent studies have considered multiple plausible root causes. For example, MP-Bench \cite{mpbench} argues that failure attribution can be inherently ambiguous, making single-label deterministic evaluation potentially underestimate attribution methods. However, such ambiguity need not be intrinsic to failure attribution. Following Who\&When, the root cause can be defined as the earliest causally responsible step whose correction would allow subsequent execution to proceed normally. From this perspective, multiple root causes may largely reflect the difficulty of obtaining precise causal annotations in complex agent systems. Thus, multi-root-cause datasets can also be viewed as a practical accommodation to annotation uncertainty, rather than evidence that failures inherently have multiple root causes.}




\section{Discussion}
\label{futuredirections}

\textbf{Timely and Proactive AgentOps.} \redd{Current AgentOps techniques are still largely reactive, with many anomalies detected only after they have propagated through long execution trajectories or caused task failures. Future AgentOps should therefore shift from post-hoc analysis toward online and proactive operations. This requires continuous anomaly detection during execution, efficient diagnosis, and timely intervention before erroneous states accumulate. For example, root cause localization may adopt a coarse-to-fine strategy, where lightweight models first narrow down suspicious regions and LLMs subsequently perform fine-grained attribution. Combined with checkpointing and in-execution correction, such mechanisms can substantially shorten the delay between anomaly occurrence, diagnosis, and recovery.}

\textbf{Accurate and Evidence-Grounded AgentOps.} \redd{Reliable AgentOps depends fundamentally on the quality of operational evidence. Conventional metrics, logs, and traces often capture structural execution behaviors but are insufficient for revealing semantic failures in reasoning, decision making, or agent interactions. Future systems should incorporate richer and higher-fidelity signals, including model-level information, execution states, memory, checkpoints, tool interactions, and inter-agent communications. More importantly, these raw observations should be transformed into semantically meaningful and risk-aware signals, such as uncertainty estimates or consistency indicators. By jointly exploiting heterogeneous evidence across agent and system levels, AgentOps can support more accurate anomaly detection, root cause localization, and resolution.}

\section{Literature Search Methodology}
\label{search}



\redd{We conducted a systematic literature search primarily using Google Scholar, complemented by backward and forward citation tracing and AI-assisted retrieval. The search covered literature published up to August 2026 and used combinations of keywords such as ``agent failure,'' ``agent error,'' ``failure attribution,'' ``anomaly detection,'' ``root cause localization,'' ``monitoring,'' and ``resolution,'' together with terms related to LLM-based agents and multi-agent systems.}

\redd{We included studies that investigate failures, operational mechanisms, evaluation benchmarks, or software-engineering techniques directly applicable or transferable to AgentOps. Studies unrelated to agent-system reliability and operations, or without a clear connection to the AgentOps lifecycle, were excluded. Given the rapid evolution of the field, our survey may still be subject to selection bias and temporal-coverage bias: relevant studies may be missed by the adopted search strategy, and new work published after August 2026 is not covered.}

\section{Conclusion}
\label{CONCLUSION}



\redd{The continuous advancement of LLM reasoning capabilities has accelerated the development of agent systems, while also introducing new operational challenges arising from their stochastic and dynamic behaviors. This paper provides a systematic definition and taxonomy of agent-system anomalies, categorizing them into intra-agent and inter-agent anomalies, and proposes AgentOps as a comprehensive operational framework encompassing monitoring, anomaly detection, root cause localization, and resolution. Looking forward, future AgentOps should move toward more proactive and evidence-grounded operations by enabling timely online intervention and leveraging richer operational signals for reliable diagnosis and recovery.}

\bibliographystyle{IEEEtran}
\bibliography{sample-base}

\end{document}